\documentclass[prd,superscriptaddress,twocolumn,showpacs,nofootinbib]{revtex4-1}

\usepackage{amsfonts}
\usepackage{amsmath}
\usepackage{amssymb}
\usepackage{upgreek}
\usepackage{bm}
\usepackage{dcolumn}
\usepackage{graphicx}
\usepackage[latin1]{inputenc}
\usepackage{latexsym}
\usepackage{rotating}
\usepackage{hyperref}
\usepackage{enumerate}
\usepackage{xspace} 
\usepackage[usenames,dvipsnames]{color}

\usepackage{ulem}
\definecolor{darkgreen}{rgb}{0.2,0.7,0.2}

\newcommand\be{\begin{equation}}
\newcommand\ba{\begin{eqnarray}}
\newcommand\ee{\end{equation}}
\newcommand\ea{\end{eqnarray}}

\newcommand{\GR}{{\mbox{\tiny GR}}}

\begin{document}
\title{Stealth Bias in Gravitational-Wave Parameter Estimation}

\author{Michele Vallisneri}
\affiliation{Jet Propulsion Laboratory, California Institute of Technology, Pasadena, CA 91109, USA.}

\author{Nicol\'as Yunes}
\affiliation{Department of Physics, Montana State University, Bozeman, MT 59717, USA.}

\date{\today}

\begin{abstract}
Inspiraling binaries of compact objects are primary targets for current and future gravitational-wave observatories. Waveforms computed in General Relativity are used to search for these sources, and will probably be used to extract source parameters from detected signals. However, if a different theory of gravity happens to be correct in the strong-field regime, source-parameter estimation may be affected by a \emph{fundamental bias}: that is, by systematic errors induced due to the use of waveforms derived in the incorrect theory. If the deviations from General Relativity are not large enough to be detectable on their own and yet these systematic errors remain significant (i.e., larger than the statistical uncertainties in parameter estimation), fundamental bias cannot be corrected in a single observation, and becomes {\emph{stealth bias}}. In this article we develop a scheme to determine in which cases stealth bias could be present in gravitational-wave astronomy. For a given observation, the answer depends on the detection signal-to-noise ratio and on the strength of the modified-gravity correction. As an example, we study three representative stellar-mass binary systems that will be detectable with second-generation ground-based observatories. We find that significant systematic bias can occur whether or not modified gravity can be positively detected, for correction strengths that are not currently excluded by any other experiment. Thus, stealth bias may be a generic feature of gravitational-wave detections, and it should be considered and characterized, using expanded models such as the parametrized post-Einstein framework, when interpreting the results of parameter-estimation analyses.
\end{abstract}

\maketitle

\section{Introduction}

Although General Relativity (GR) has been shown to be an excellent description of Nature in all astronomical, astrophysical, and laboratory observations carried out so far \cite{2006LRR.....9....3W}, this theory remains largely untested in the nonlinear and dynamical {\emph{strong-field}} regime, where gravitational fields are large and rapidly evolving, and velocities are not small. Future gravitational-wave (GW) observations of the late inspiral and coalescence of compact binaries will probe this regime and allow for new tests of GR. To do so, however, the assumption that GR is correct must be relaxed in GW science, and data analysis must be carried out with more general waveform families than those predicted by GR.

The parameterized post-Einsteinian (ppE) framework~\cite{2009PhRvD..80l2003Y} was recently proposed as a generic solution for this need. In this framework, one deforms the GR waveforms  through amplitude ($A$) and phase ($\Psi$) corrections expanded in a polynomial basis: $\delta \Psi = \beta_{i} u^{b_{i}}$ and $\delta A = \alpha_{i} u^{a_{i}}$, where $u = \pi M f$ is the reduced GW frequency and $(\alpha_{i},\beta_{i},a_{i},b_{i})$ are ppE parameters. Of these, $a_{i}$ and $b_{i}$ control the type of modified-gravity (MG) corrections, while $\alpha_{i}$ and $\beta_{i}$ control their magnitude. Recent data-analysis investigations of this framework, both in its full form~\cite{2011PhRvD..84f2003C} and in a reduced version restricting the allowed $b_{i}$ to those that appear in post-Newtonian (PN) expansions \cite{2012PhRvD..85h2003L}, suggest that it could be used in realistic settings.

The simplest ppE model includes a single amplitude and phase correction to the waveform, and it is still sufficiently general to extend the PN-coefficient tests\footnote{These tests treat the coefficients in the PN expansion of the inspiral-waveform phasing as independent parameters rather than fixed functions of the binary parameters, and verify the consistency of their measured values.} of Refs.\ \cite{Arun:2006hn,Arun:2006yw,2010PhRvD..82f4010M} to all currently studied MG theories, including Brans--Dicke theory~\citep{Will:1994fb,Will:2004xi,Berti:2005qd,2009PhRvD..80d4002S,2009CQGra..26o5002A,Keppel:2010qu,2010PhRvD..81f4008Y,Yunes:2011aa,Healy:2011ef}, dynamical Chern--Simons gravity~\citep{Sopuerta:2009iy,Yunes:2009hc,2012PhRvD..85f4022Y,Yagi:2012ya,Yagi:2012vf}, phenomenological massive-graviton propagation~\citep{Will:1997bb,Scharre:2001hn,Will:2004xi,Berti:2005qd,2010PhRvD..81f4008Y,2011PhRvD..84j1501B}, gravitational Lorentz violation~\citep{2012PhRvD..85b4041M}, gravitational parity violation~\citep{Yunes:2010yf,2012PhRvD..85f4022Y}, violations of Local Position Invariance~\citep{Yunes:2009bv}, and the existence of extra-dimensions~\citep{Yagi:2011yu}. More complicated ppE models posit a piecewise-specified hybrid waveform model that includes the merger and ringdown phases, as well as multi-exponent inspiral corrections \cite{2009PhRvD..80l2003Y}.

Neglecting the prospect that GR might be an inaccurate description of Nature in the strong-field regime may lead to the mischaracterization of the GW Universe once detections are made. This pitfall has been labeled \emph{fundamental bias}~\cite{2009PhRvD..80l2003Y} in analogy to other types of bias in observational astronomy. Imagine, for example, that the behavior of gravitation is correctly described by a MG theory that allows for scalar radiation only in the very late stages of inspiral. If so, the inspiral will proceed faster than would be expected in GR, and less power may be emitted in GWs during ringdown. Such events might still be detected with GR templates, albeit with suboptimal efficiency, but they will appear closer than they really are, introducing a systematic error in the estimated luminosity distance.
 
This paper discusses how we may quantify the degree to which fundamental bias could lead us to incorrect inferences about the parameters of detected GW sources. The magnitude of the errors introduced will of course depend on the magnitude of the MG correction. In this first study, we will model the correction with the simplest ppE inspiral waveform, which includes a single-exponent modification to the phasing. We ask the following question: As a function of correction exponent $b$ and magnitude $\beta$, what is the detection signal-to-noise ratio such that the systematic error due to using GR templates is equal to the statistical, noise-induced error in the determination of the source parameters, such as the chirp mass and mass ratio? We then ask: At that SNR, is there sufficient statistical evidence, as quantified by Bayesian model comparison \cite{2005blda.book.....G}, to claim that the MG model is preferred? If the systematic error is relatively large, but model comparison still prefers GR, we would found ourselves in a situation of \emph{stealth bias} \cite{2011PhRvD..84f2003C} that makes it logically impossible to correct the systematic effects of MG theories.

Fundamental bias in GW observations is similar to the \emph{mismodeling bias} considered by Cutler and Vallisneri (who call it \emph{theoretical} error \cite{Cutler:2007mi}): the idea is that the approximations used to solve the Einstein equations and construct waveform templates may introduce parameter-estimation errors that are larger than statistical uncertainties. Cutler and Vallisneri developed simple formulas to estimate the likely mismodeling error; in this paper we use them in the ppE context. We also use Vallisneri's recent formula \cite{2012PhRvD..86h2001V} to compute the Bayesian odds ratio, as measured in GW observations, for GR versus a MG theory linked to GR by one or more continuous parameters. These formulas are valid for sufficiently large detection SNRs. For weaker signals, a full-scale Monte Carlo approach is needed \cite{2011PhRvD..84f2003C}. In this paper, we limit ourselves to exemplifying the logic of our approach with simpler analytic tools of more limited applicability, and we leave a detailed Monte Carlo analysis for future work.

As an example, we concentrate on the signals from the circular--adiabatic inspiral of nonspinning stellar-mass binaries, as detected by a network of second-generation ground-based observatories. The signals include ppE phasing corrections with a range of exponents and strengths. We find that, for likely detection SNRs, three cases are indeed possible, depending on the strength of the correction: 
\begin{enumerate}
\item[i.] Fundamental bias is significant, but statistical evidence is sufficient to positively detect MG effects, so the bias can be corrected; 
\item[ii.] Fundamental bias is significant, but statistical evidence is \emph{insufficient} to detect MG effects, so stealth bias is present; 
\item[iii.] Fundamental bias is not significant and MG effects cannot be detected.
\end{enumerate}
For exponents $b$ corresponding to quadrupole- or higher-order PN terms (i.e., $b \geq -5/3$), stealth bias can occur for correction strengths $\beta$ that are \emph{not} currently excluded by other constraints, such as those from binary pulsars.

This paper is organized as follows. Section \ref{sec:bias} describes our formulation; Sec.\ \ref{sec:results} discusses our analysis and results; Sec.\ provides are conclusions. Throughout this paper we use the conventions of Misner, Thorne, and Wheeler \cite{Misner:1973cw}, such as geometric units with $G=c=1$. 

\section{Formulation}
\label{sec:bias}

In this section we describe our analytical formulation. Subsection \ref{sec:ppe} introduces ppE waveforms for inspiraling binaries; Subsecs.\ \ref{sec:systematic} and \ref{sec:detect} provide details about the estimation of fundamental bias and the model-comparison detection of MG; Subsec.\ \ref{sec:stealth} ties all of these together to characterize the impact of fundamental bias in typical detections and the occurrence of stealth bias.
 
\subsection{Systematic bias and the ppE framework}
\label{sec:ppe}

GW searches performed with matched filtering can be affected by systematic bias if the theoretical signal templates used in the search do not match exactly the true signals that appear in the data. Symbolically, $h_\mathrm{true} = h_\mathrm{theory} + \delta h$, where the unmodeled correction $\delta h$ can have different origins~\cite{Yunes:2011ws}:
\begin{itemize}
\item \textbf{Mismodeling}: caused by the approximations used in solving the equations that describe the motion of the GW source, and the generation and propagation of GWs~\cite{Cutler:2007mi}.
\item \textbf{Instrumental}: caused by the approximations used in modeling the response of the detector to GWs~\cite{Vitale:2011wu,Whelan:2012ur}.\footnote{Of course, the first thing that comes to mind when considering instrumental effects is the parameter-estimation error due to detector noise, which has a stochastic rather than systematic nature. These errors \emph{may} be difficult to characterize statistically if the noise is significantly non-Gaussian or nonstationary, but they are distinct from the systematic effects that we address in this paper, since they can be alleviated only by improving the instrument itself, rather than its theoretical description or the modeling of sources.}
\item \textbf{Astrophysical}: caused by unmodeled astrophysics (e.g., neutron-star hydrodynamics in NS--NS inspirals) or objects (e.g., a third star in the close vicinity of a binary)~\cite{Yunes:2010sm,Yunes:2011ws,Kocsis:2011dr}.
\item \textbf{Fundamental}: caused by MG corrections to Einstein's equations~\cite{2009PhRvD..80l2003Y,2011PhRvD..84f2003C,2012PhRvD..86h2001V}.
\end{itemize}

Mismodeling bias is remedied by deriving ever more accurate solutions to the field equations: for binary inspirals, this goal is currently pursued by pushing the PN approximation to higher orders, and by integrating together PN and numerical-relativity results with analytical resummation and fitting techniques, such as the effective-one-body scheme.
Instrumental bias is reduced by careful detector modeling and characterization.
Astrophysical bias is expected to be irrelevant for most ground-based binary observations; but even if this were not the case, astrophysical effects should present themselves differently (or not at all) in observations of different systems, whereas fundamental bias, if present, would appear equally in all observed systems.

In this paper, we concentrate on the inspiral signals from compact-binary coalescences. We consider inspirals that are circular and adiabatic, with negligible spin effects, and neglect mismodeling bias by assuming the GW emission is well-described by the restricted PN waveform in the frequency-domain, stationary-phase approximation~\cite{1994PhRvD..49.2658C,Flanagan:1997sx,Flanagan:1997kp,Droz:1999qx}. In GR, the resulting signals can be written as
\be
\label{eq:spa}
h_{\GR}(f) = A_{\GR}(f)  e^{i \Psi_{\GR}(f)}\,,
\ee
where $A_{\GR}(f) = \mathcal{A} \, u^{-7/6} \left[ 1 + \cdots\right]$ (we neglect PN amplitude corrections, symbolized with ellipses in the above equation), $u = \pi {\cal{M}} f$ is the reduced frequency, ${\cal{M}} = \eta^{3/5} M$ the chirp mass, $\eta = m_{1} m_{2}/M^{2}$ the symmetric mass ratio, $M = m_{1} + m_{2}$ the total mass, and $f$ the GW frequency. The constant amplitude ${\cal{A}}$ depends on the chirp mass, the luminosity distance, and the detector's antenna patterns~\cite{1994PhRvD..49.2658C,Flanagan:1997sx,Flanagan:1997kp,Droz:1999qx}. The quantity $\Psi_{\GR}(f)$ in Eq.\ \eqref{eq:spa} is the GW phase, given in the PN approximation by
\begin{align}
\Psi_{\GR} &= 2 \pi f t_{c}  - \phi_{c} - \frac{\pi}{4} 
\nonumber \\
& + \frac{3 u^{-5/3}}{128} \biggl\{1 + \sum_{k=2}^{7} \Bigl[\psi_{k} + \frac{1}{3} \bar{\psi}_{k} \log(u) \Bigr] \eta^{-k/5} u^{k/3} \biggr\},
\label{eq:GRphase-PN}
\end{align}
where the constant coefficients $(\psi_{k},\bar{\psi}_{k})$ can be found (for instance) in Ref.\ \cite{Cutler:2007mi}.

Under these assumptions, the unmodeled corrections enumerated above can be represented by a continuous and (in principle) predictable deformation of the GW phase $\Psi(f)$ and amplitude $A(f)$. The particular deformation depends on the systematic effect. For mismodeling, we expect corrections within the structure of the PN series ($\delta A \propto f^{-7/6 + k_{A}/3}$ for the amplitude and $\delta \Psi \propto f^{-5/3 + k_{\phi}/3}$ for the phase, with integers $k_{A},k_{\phi} > 0$), beyond the highest known perturbative order (i.e., $k_{A} > 5$ for the amplitude and $k_{\phi} > 7$ for the phase).

For astrophysical effects, we expect corrections to arise almost always with ``negative'' PN exponents \cite{Yunes:2009yz,Yunes:2010sm,Yunes:2011ws,Kocsis:2011dr}. For example, an accretion disk \cite{Yunes:2009yz}, the presence of a third body \cite{Yunes:2010sm}, and orbital eccentricity \cite{Yunes:2009yz} all introduce GW phase corrections $\delta \Psi \propto f^{-5/3 -k'_\phi}$, with integer $k'_\phi > 0$. Physically, this frequency-dependence corresponds to astrophysical effects becoming less important for tighter binaries, where strong-field effects become dominant. 

Moving on to unmodeled corrections originating from fundamental physics,
the amplitude and phase deformations $\delta A$ and $\delta \Psi$ can always be expressed as sums of frequency powers, provided that $\delta A$ and $\delta \Psi$ remain \emph{analytic} at all frequencies sampled during the inspiral:
\be
\delta A = A_{\GR}(f) \sum_{k=1}^K \alpha_k u^{a_{k}}\,,  
\qquad
\delta \Psi(f) = \sum_{k=1}^{K} \beta_k u^{b_{k}}\,,  
\label{Gen-Ppe}
\ee
where $(\alpha_k, \beta_k, a_k, b_k) \in \mathbb{R}$ for all $k$, and where we have included the $A_{\GR}$ prefactor in $\delta A$.
We have neglected possible logarithmic terms for simplicity, but they can be included easily in the same fashion. This ppE model introduces $4 K$ new parameters in the waveform; the simplest version of this model would allow only a single exponent:
\be
\delta A = A_{\GR}(f) \, \alpha \, u^{a}\,,
\qquad
\delta \Psi(f) = \beta \, u^{b}\,.
\ee
Indeed, it can be shown that such a parametric deformation is sufficiently general to model all known MG corrections to the waveform to leading PN order \cite{2009PhRvD..80l2003Y,2011PhRvD..84f2003C}, provided that the two tensor polarizations are dominant, as in GR. Otherwise, a second term would be needed in the phase and amplitude \cite{Arun:2012hf,Chatziioannou:2012rf}. 

Furthermore, a convincing argument can be made that $a$ and $b$ should be restricted to a few discrete values \cite{Chatziioannou:2012rf}. Suppose that the adiabatic-inspiral waveform is derived from an energy-balance equation with modified binding energy and flux of the form
\begin{align}
E &= E_{0} \, v^{2} \left[1 + \left(\cdots\right)_\mathrm{PN} + \delta E \, v^{k} \right]\,,
\\
\dot{E} &= \dot{E}_{0} \, v^{10} \left[1 + \left(\cdots\right)_\mathrm{PN} + \delta \dot{E} \, v^{m} \right]\,,
\end{align}
where $v$ is the relative velocity of the binary components, ellipses stand for higher-order PN terms, and $E_{0},\dot{E}_{0},\delta E$ and $\delta\dot{E}$ are all constants that may depend on the source parameters and on the MG coupling constants. The exponents $k$ and $m$ must be integers, otherwise $E$ or $\dot{E}$ would not be analytic, and we would lose the guarantee that the equations have a unique solution of hyperbolic character by the Picard--Lindel\"{o}f theorem\footnote{Given the differential equation $dy/dt = f(t,y(t))$, with initial value $y(t_{0}) = y_{0}$, a unique solution exists for all $t \in (t_{0} - \epsilon,t_{0} + \epsilon)$ provided that $f$ is Lipschitz continuous in $y$ and continuous in $t$. A noninteger value of $k$ and $m$ would lead to a differential equation with a non-Lipschitz continuous source term, with possible loss of uniqueness.}. Furthermore, we must have $k \geq -2$ and $m \geq -10$, otherwise $E$ and $\dot{E}$ would not reduce to the GR result in the weak-field limit.
These constraints lead to the deformations
\be
\delta A = A_{\GR}(f) \, \alpha \, u^{\bar{a}/3}\,,  
\qquad
\delta \Psi(f) = \beta \, u^{\bar{b}/3}\,,  
\ee
where $(\bar{a},\bar{b}) \in {\mathbb{Z}}$, with $({\bar{a}},{\bar{b}})> (-10,-15)$ \cite{Chatziioannou:2012rf}. 

In this paper we concentrate on phasing corrections by setting $\alpha = 0$ and choosing $\bar{b} \in \{-7, -6, -5, -4, -3, -2, -1, 1, 2\}$. Different values of $\bar{b}$ represent different types of MG effects: $\bar{b} = -7$ corresponds to the leading--PN-order correction in Brans--Dicke theory~\cite{Will:1994fb,Scharre:2001hn,Will:2004xi,Berti:2004bd,2010PhRvD..81f4008Y,Yunes:2011aa,Healy:2011ef} or in Einstein--dilaton--Gauss--Bonnet gravity~\cite{Yunes:2011we,2012PhRvD..85f4022Y}; $\bar{b} = -3$ to the leading-order term in a phenomenological massive graviton theory~\cite{Will:1997bb,Will:2004xi,Berti:2004bd,2009CQGra..26o5002A,2009PhRvD..80d4002S,2010PhRvD..81f4008Y,2009PhRvD..80d4002S,2012PhRvD..85b4041M}, $\bar{b} \geq -5$ (but $\neq -4$) to the modified-PN scheme of Refs.\ \cite{Arun:2006hn,Arun:2006yw,2010PhRvD..82f4010M,2012PhRvD..85h2003L}, and $\bar{b}=-1$ to dynamical Chern--Simons gravity~\cite{Alexander:2009tp,2012PhRvD..85f4022Y,Yagi:2012vf}. Notice, in particular, that the modified-PN scheme is clearly a sub-case of the ppE scheme. We omit $\bar{b} = 0$ because the resulting correction would be degenerate with an arbitrary constant in the phase. We do not consider $\bar{b} < -7$ because the values that we study provide enough information to observe a consistent trend as $\bar{b}$ becomes more negative. Moreover, for $\bar{b} \leq -7$, binary pulsar observations can do a better job at constraining modified gravity theories than GWs observations \cite{Yunes:2010qb}. We do not consider $\bar{b} > 2$, as this would correspond to terms of higher than $3.5$ PN order, which we do  not account for in the $\Psi_{\GR}$.

\subsection{Quantifying the bias}
\label{sec:systematic}

Let us assume that a GW detection is reported for a dataset $s$ that contains the waveform
\begin{equation}
h_\mathrm{full}(\vec{\theta}_{\rm tr}) = h(\vec{\theta}_{\rm tr}) + \delta h(\vec{\theta}_{\rm tr}),
\end{equation}
where $\vec{\theta}_\mathrm{tr}$ is the vector of parameters that describes the GW source and source--detector geometry, $h$ is the approximated waveform family used to filter the data, and $\delta h$ is the unmodeled correction to $h$.
Following Cutler and Vallisneri \cite{Cutler:2007mi}, we compute the \emph{theoretical} error $\delta \vec{\theta}_\mathrm{th}$ induced by matched-filtering with $h$ instead of $h_\mathrm{full}$.

The theoretical error $\delta \vec{\theta}_\mathrm{th}$ is defined as the displacement $\vec{\theta}_\mathrm{bf} - \vec{\theta}_\mathrm{tr}$ between the true parameters $\vec{\theta}_\mathrm{tr}$ and the \emph{best-fit} parameters $\vec{\theta}_\mathrm{bf}$ that would maximize the likelihood in the absence of noise. 
When $\delta h(\vec{\theta}_\mathrm{tr})$ is negligibly small, $\vec{\theta}_\mathrm{bf} = \vec{\theta}_\mathrm{tr}$; as $\delta h(\vec{\theta}_\mathrm{tr})$ grows in magnitude, $\vec{\theta}_\mathrm{bf}$ is displaced further and further away along the parameter-space direction in which $h(\vec{\theta}_\mathrm{bf})$ can reproduce $h(\vec{\theta}_\mathrm{tr}) + \delta h(\vec{\theta}_\mathrm{tr})$ most closely.

To leading order in $\delta h$, $\delta \vec{\theta}_\mathrm{th}$ is given by \cite{Cutler:2007mi}
\begin{equation}
\label{eq:bias}
\delta \vec{\theta}_\mathrm{th}
= (F_\mathrm{bf}^{-1})^{\alpha \beta}
\bigl(h_{,\beta}(\vec{\theta}_\mathrm{bf}) \big| \delta h(\vec{\theta}_\mathrm{bf}) \bigr),
\end{equation}
where $h_{,\beta} = \partial h/\partial \theta^\beta$ are partial derivatives of the waveform with respect to source parameters, $F_{\alpha \beta} = (h_{,\alpha}|h_{,\beta})$ is the \emph{Fisher matrix}, here evaluated at $\theta_\mathrm{bf}$, and 
\begin{equation}
(g_1|g_2) = 4 \, \mathrm{Re} \int_0^\infty \frac{\tilde{g_1}^*(f) \tilde{g_2}(f)}{S_n(f)} \, \mathrm{d}f,
\end{equation}
is the noise-weighted signal inner product, with $S_n(f)$ the one-sided power spectral density of detector noise (see, e.g., \cite{1994PhRvD..49.2658C}). The inner product defines a signal \emph{norm} $|h|$ by way of $|h|^2 = (h|h)$.

In Eq.\ \eqref{eq:bias}, the waveform correction $\delta h$ is projected onto the waveform derivatives, and the projection cosines are mapped into parameter errors by the inverse Fisher matrix $F^{-1}$, thus taking into account possible parameter covariances. Note that the resulting $\delta \vec{\theta}_\mathrm{th}$ is independent of the detection SNR, since both $F$ and $(h_{,\beta}(\theta_\mathrm{bf})|\delta h(\theta_\mathrm{bf}))$ are quadratic in the waveform amplitude.

Equation \eqref{eq:bias} is only accurate for small $\delta \vec{\theta}_\mathrm{th}$---more precisely, for perturbations small enough that $h(\vec{\theta}_\mathrm{bf} - \delta \vec{\theta}_\mathrm{th}) \simeq h(\vec{\theta}_\mathrm{bf}) - h_{,\alpha} \delta \theta^\alpha_\mathrm{th}$. Cutler and Vallisneri \cite{Cutler:2007mi} discuss more sophisticated versions of Eq.\ \eqref{eq:bias} that can be applied to larger perturbations, but in this paper we adopt the simpler Eq.\ \eqref{eq:bias}, not least because the other ingredients in our formulation depend on $\delta h(\vec{\theta}_\mathrm{tr})$ being small.

\subsection{Detecting modified gravity}
\label{sec:detect}

Following Vallisneri \cite{2012PhRvD..86h2001V} (see also~\cite{2011PhRvD..84f2003C}), we define a MG correction $\delta h$ to the signal $h$ to be detectable when the \emph{odds ratio} of the Bayesian evidences for the MG and pure-GR scenarios, used as a detection statistic, is large enough that the \emph{false-alarm probability} of favoring the MG hypothesis when GR is in fact correct is suitably small. More precisely, we compute the odds ratio
\newcommand{\Oc}{\mathcal{O}}
\newcommand{\ag}{\mathrm{MG}}
\newcommand{\gr}{\mathrm{GR}}
\begin{equation}
\label{eq:oddsratio}
\Oc = \frac{P(\ag|s)}{P(\gr|s)} = \frac{
P(\ag) \int p(s|\vec{\theta},\vec{\lambda}) \, p(\vec{\theta},\vec{\lambda}) \, \mathrm{d}^{k}{\theta} \, \mathrm{d}^{m} {\lambda}
}{
P(\gr) \int p(s|\vec{\theta}) \, p(\vec{\theta}) \, \mathrm{d}^{k}{\theta}},
\end{equation}
where $P(\ag)$ and $P(\gr)$ are the prior probabilities that MG and GR are correct, $p(s|\vec{\theta},\vec{\lambda})$ is the likelihood that the detector data $s$ contains the MG waveform $h(\vec{\theta}) + \delta h(\vec{\theta},\vec{\lambda})$, $p(s|\vec{\theta})$ is the likelihood that $s$ contains the pure-GR waveform $h(\vec{\theta})$, and $p(\vec{\theta},\vec{\lambda}) = p(\vec{\theta}) p(\vec{\lambda})$ and $p(\vec{\theta})$ are the prior probability densities for the source parameters $\vec{\theta}$ and the MG parameters $\vec{\lambda}$.

If the true signal is MG, using MG templates would improve the fit to the data and increase the maximum value attained by the MG likelihood relative to the GR likelihood. On the other hand, the evidence for the more complicated, higher-dimensional MG model is reduced by the smaller \emph{prior mass} within the support of the likelihood---the mechanism by which Bayesian inference embodies Occam's principle of parsimony. As signals get stronger, the improvement in the likelihood grows exponentially with the (squared) detection SNR, and eventually it overcomes the effect of the priors.

Even if we fix the true signal, the odds ratio $\Oc$ remains a random variable, because it depends on the realization of detector noise, by way of the likelihoods. For sufficiently large detection SNR, it can be shown \cite{2012PhRvD..86h2001V} that Eq.\ \eqref{eq:oddsratio} becomes remarkably simple: for the cases when the underlying signal is pure-GR or MG respectively, we find that
\newcommand{\snr}{\mathrm{SNR}}
\newcommand{\ff}{\mathrm{FF}}
\begin{equation}
\label{eq:oc}
\begin{aligned}
\Oc_\gr &\propto e^{x^2/2}, \\
\Oc_\ag &\propto e^{x^2/2 + \sqrt{2} \, x \, \snr_\ag + \snr_\ag^2}.
\end{aligned}
\end{equation}
In this equation:
\begin{itemize}
\item The dependence on the noise realization enters exclusively through $x$, a normal random variable with zero mean and unit variance. Technically, $x$ is given by the inner product of detector noise with the ``MG-unique'' component of the MG correction $\delta h$ (the component that is orthogonal to the derivatives of the signal with respect to the GR parameters).
\item The constant of proportionality is the same in both rows: it is a function of $P(\ag)$, $P(\gr)$, and of the estimation errors and prior-density widths for the MG parameters $\vec{\lambda}$. As we shall see below, the fact that this constant is the same under both hypotheses allows us to disregard it when we analyze the statistics of our detection scheme.
\item Last, $\snr_\ag \equiv \snr \sqrt{1 - \ff}$ is (by definition) the \emph{effective MG-detection SNR}, with $\ff$ the \emph{fitting factor} between the pure-GR and MG waveforms,
\begin{equation}
\label{eq:ff}
\ff({\vec{\theta}},\vec{\lambda}) = \max_{{\vec{\theta}}'} \frac{\bigl( h({\vec{\theta}}')\big| h({\vec{\theta}}) + \delta h({\vec{\theta}},\vec{\lambda}) \bigr)
}{
|h({\vec{\theta}}')| \times |h({\vec{\theta}}) + \delta h({\vec{\theta}},\vec{\lambda})|
}\,.
\end{equation} 
This FF describes the maximum fraction of detection SNR that can be recovered when searching for MG signals using GR templates.
The FF is itself independent of SNR, and in our application it is given by
\begin{equation}
\frac{\bigl( h(\vec{\theta}_\mathrm{tr} + \delta {\vec{\theta}}_\mathrm{th})\big| h(\vec{\theta}_\mathrm{tr}) + \delta h(\vec{\theta}_\mathrm{tr},\vec{\lambda}) \bigr)
}{
|h(\vec{\theta}_\mathrm{tr} + \delta {\vec{\theta}}_\mathrm{th})| \times |h({\vec{\theta}_\mathrm{tr}}) + \delta h({\vec{\theta}_\mathrm{tr}},\vec{\lambda})|
}\,,
\end{equation} 
or also by $1 - \frac{1}{2} |\delta h({\vec{\theta}}_\mathrm{tr},\vec{\lambda})|^2/|h({\vec{\theta}}_\mathrm{tr})|^2$ \cite{2012PhRvD..86h2001V}. These formulas are equivalent, but only to leading order in $\vec{\lambda}$.
\end{itemize}

By solving both rows of Eq.\ \eqref{eq:oc} for $x$ as a function of $\Oc$, we can derive the cumulative probability distributions of $\Oc_\gr$ and $\Oc_\ag$ from the normal probability distribution $p(x) = e^{-x^2/2} / \sqrt{2\pi}$. For instance,
\begin{equation}
\label{eq:pf}
P(\Oc_\gr > \Oc^*) = \int_{x(\Oc^*)}^{\infty} p(x) \, \mathrm{d}x:
\end{equation}
this is exactly the false-alarm probability $P_F$ in a \emph{decision scheme} that declares an MG detection when the odds ratio $\Oc$ computed from the data is larger than the \emph{threshold} $\Oc^*$. By contrast, the true-detection probability (also known as the detection \emph{efficiency}) is given by $P_E \equiv P(\Oc_\ag > \Oc^*)$, which we compute along the lines of Eq.\ \eqref{eq:pf}, and which is a function of $\mathrm{SNR}_\mathrm{MG}$. In practice, we set $\Oc^*$ by requiring a sufficiently small $P_F$. In this paper, we use $P_F = 10^{-4}$, which seems appropriate for the tens of detections expected from second-generation ground-based detectors (but see also the discussion in Sec.\ III of \cite{2012PhRvD..86h2001V}).

We can now replace the vague statement that began this section with a precise quantitative condition: we will say that a MG correction becomes detectable once $\mathrm{SNR}_\mathrm{MG}$ is large enough to yield an efficiency of 1/2. This is a conventional value, but once $P_E$ reaches 1/2 it grows rapidly to 1 with increasing $\mathrm{SNR}_\mathrm{MG}$. For $P_F = 10^{-4}$, we get $P_E = 1/2$ when $\mathrm{SNR}_\mathrm{MG} = 2.75$. Because $\snr_\ag \equiv \snr \sqrt{1 - \ff}$, the SNR at which MG effects are detectable is then simply $\mathrm{SNR}^\mathrm{detect} = 2.75 (1 - \ff)^{-1/2}$. We thus confirm our expectation that MG detectability improves for larger detection SNR or for larger MG corrections, which lead to smaller FF.

\subsection{Statistical error and stealth bias}
\label{sec:stealth}

We have by now collected the tools to estimate the fundamental error $\delta {\vec{\theta}}_\mathrm{th}$ due to MG corrections, and to determine the detection SNR necessary to positively detect the presence of MG corrections of a certain magnitude.
The last ingredient in our formulation is the statistical error in the determination of the source parameters ${\vec{\theta}}$, which we estimate as the (square-root) diagonal elements of the inverse Fisher matrix, computed at ${\vec{\theta}}_\mathrm{tr}$ for GR waveforms:
\begin{equation}
\delta {{\theta}}_\mathrm{stat}^\alpha = \sqrt{(F_\mathrm{tr}^{-1})^{\alpha\alpha}}
\end{equation}
(no summation is implied by the repeated index). This expression implies that $\delta {{\theta}}_\mathrm{stat}^\alpha$ scales as $1/\mathrm{SNR}$. Indeed, to leading order in $1/\mathrm{SNR}$, the inverse Fisher matrix describes the variance of the maximum-likelihood parameter estimator across noise realizations, as well as the shape of the Bayesian posterior parameter distribution around its mode when priors can be neglected \cite{2008PhRvD..77d2001V}.

For a fixed MG-correction magnitude (a fixed $\beta$), the statistical error decreases with increasing detection SNR, but the systematic bias remains constant and eventually becomes the limiting factor for parameter-estimation accuracy. If this happens before MG can be positively detected, the observation may suffer from stealth bias. Thus, we will look for the detection $\mathrm{SNR}$ at which $\max_\alpha {\delta \theta}_\mathrm{th}^\alpha / \delta {\theta}_\mathrm{stat}^\alpha = 1$, 
and compare it to the detection SNR at which $\mathrm{SNR}_\mathrm{MG} \equiv \mathrm{SNR} \sqrt{1 - \mathrm{FF}} = 2.75$. We will denote these two limiting SNRs as $\mathrm{SNR}^\mathrm{bias}$ and $\mathrm{SNR}^\mathrm{detect}$, respectively.

We should not forget that our formulation is only valid for sufficiently large SNRs. It is difficult in general to determine if higher-order contributions change an analytical result without actually computing them, but we can at least check that the leading-order result is self-consistent. Indeed,  we adopt Vallisneri's 2008 criterion \cite{2008PhRvD..77d2001V} for the statistical errors predicted by the Fisher-matrix formalism, and we require that the error made in approximating $h({\vec{\theta}}_\mathrm{tr} + \delta {\vec{\theta}}_\mathrm{stat}) - h({\vec{\theta}}_\mathrm{tr})$ as $h_{,\alpha} \delta {\theta}^\alpha_\mathrm{stat}$ be sufficiently small (0.1 in norm) on most (95\%) of the $1\mbox{--}\sigma$ error surface described by $(F^{-1})^{\alpha\beta}$. For the theoretical errors, we require that the FFs computed in the two ways described below Eq.\ \eqref{eq:ff} be consistent to $1\%$.

\section{Analysis and results}
\label{sec:results}

We examine three representative binary GW sources for second-generation interferometric detectors such as Advanced LIGO \cite{2010CQGra..27h4006H}: neutron-star--neutron-star binaries with $(1.4 + 1.4)M_\odot$ component masses; neutron-star--black-hole binaries with $(1.4 + 5)M_\odot$ masses; and black-hole--black-hole binaries with $(5 + 10)M_\odot$ masses.

We concentrate on the inspiral phase of coalescence, which we model as quadrupolar and adiabatically quasi-circular with 3.5PN-accurate phasing. We truncate the waves at the innermost stable circular orbit of a point-particle in a Schwarzschild background (assuming GR), and we neglect spin effects and PN amplitude corrections. The resulting waveforms are described by nine parameters: the two masses (or the chirp and reduced masses), the time and inspiral phase at coalescence, two sky-position angles, two angles that describe the binary inclination and GW polarization, and the luminosity distance (see \cite{Cutler:2007mi} and \cite{2011PhRvD..84f2003C} for a similar waveform prescription). We assume a simultaneous detection by three second-generation detectors with the LIGO Hanford, LIGO Livingston, and Virgo geometries and relative delays \cite{T010110}, and with identical broadband-configuration power spectral densities, as given by Eq.\ (10) of \cite{2011PhRvD..84f2003C}. Furthermore, we assume that GW-detector noise is Gaussian and stationary, as required by the Cutler--Vallisneri \cite{Cutler:2007mi} and Vallisneri \cite{2012PhRvD..86h2001V} formalisms. 

For these systems, we consider ppE phasing corrections $\delta \Psi$ as described in Sec.\ \ref{sec:ppe}, and we compare $\mathrm{SNR}^\mathrm{bias}$ and $\mathrm{SNR}^\mathrm{detect}$ as a function of the MG-correction magnitude $\beta$ for a range of exponents $\bar{b}$. For each mass combination, each $\bar{b}$, and each $\beta$, we randomly select 1,000 configurations of the phase and angle parameters from the appropriate uniform distributions (e.g., sky positions are chosen randomly on the celestial sphere). The luminosity distance is reabsorbed in the SNR scaling, while the time of coalescence has no effect on our computation. For each configuration we compute $\mathrm{SNR}^\mathrm{detect}$ and $\mathrm{SNR}^\mathrm{bias}$, and we report their median values. 
The condition $\max_\alpha \delta {\theta}_\mathrm{th}^\alpha / \delta {\theta}_\mathrm{stat}^\alpha = 1$ that yields the latter is almost invariably satisfied first for the chirp-mass parameter. Statistical fluctuations around the median turn out to be rather small (a few percent).

The $\delta {\vec{\theta}}_\mathrm{stat}$ consistency check is satisfied for detection SNRs ranging from 10 to 100, typically $\sim 50$, but our results for lower SNRs should be at least representative of trends. The $\delta {\vec{\theta}}_\mathrm{th}$ check is satisfied for a maximum $\beta$ that depends on $\bar{b}$, and which sets the largest $\beta$ that we investigate. By contrast, the smallest $\beta$ that we study corresponds to $\min(\mathrm{SNR}^\mathrm{detect},\mathrm{SNR}^\mathrm{bias}) = 100$, a relatively large detection SNR that would be achieved very rarely in volume-limited searches \cite{2010CQGra..27q3001A}.

\begin{figure}
\includegraphics[width=0.49\textwidth]{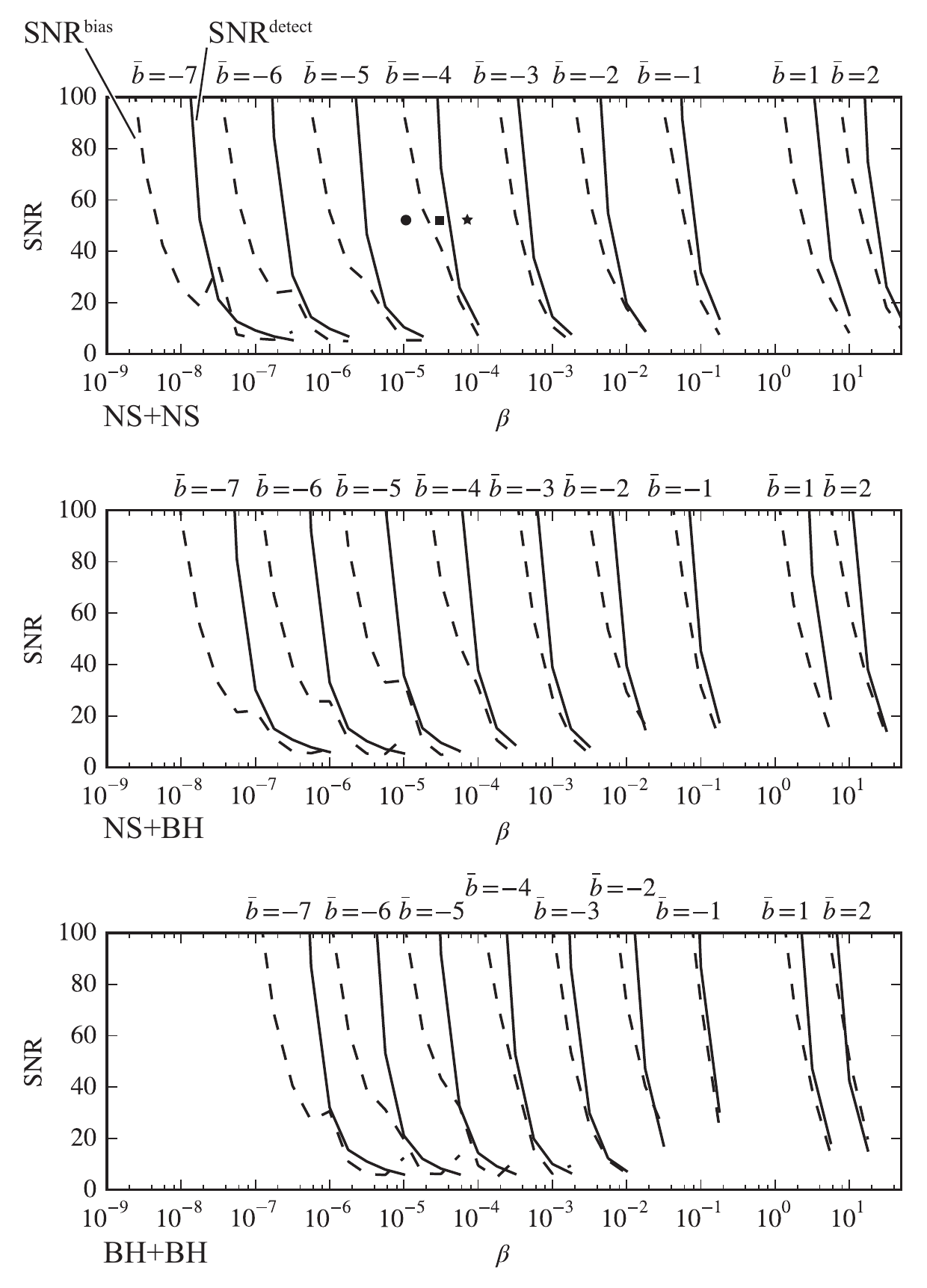}
\caption{\label{Main-Figure}
$\mathrm{SNR}^\mathrm{detect}$ (solid curves) and $\mathrm{SNR}^\mathrm{bias}$ (dashed curves) as  a function of $\beta$ for $\bar{b} = -7, -6, -5, -4, -3, -2, -1,1$ and $2$ (left to right), for a NS--NS system with $(m_1,m_{2}) =(1.4,1.4) \, M_\odot$ (top), a NS--BH system with $(m_1,m_{2}) =(1.4,5) \, M_\odot$ (mid) and a BH--BH system (bottom) with $(m_1,m_{2}) =(5,10) \, M_\odot$. The symbols in the top plot are described in the main text.}
\end{figure}

Figure~\ref{Main-Figure} presents the main results of this paper, with the solid curves plotting $\mathrm{SNR}^\mathrm{detect}$ (once again, the detection SNR above which MG can be detected positively), and the dashed curves plotting $\mathrm{SNR}^\mathrm{bias}$ (the SNR above which the largest ratio of fundamental error to statistical error reaches one). Both sets of curves are plotted as a function of the MG-correction magnitude $\beta$; the curves in each set correspond to $\bar{b} = -7, -5, -4, -3, -2, -1, 1$, and $2$, from left to right as labeled. The top, mid, and bottom panels report results for our $(1.4+1.4)M_\odot$, $(1.4+5)M_\odot$, and $(5+10)M_\odot$ systems, respectively.

This figure reveals a few interesting features. First, for the same detection SNR, more massive systems require larger $\beta$ before MG can be detected. This must happen because the larger the mass, the fewer the number of useful GW cycles in the detector's band, so the signals become relatively featureless, and higher FFs can be obtained for the same $\beta$. Of course, this feature would probably be affected if one included the merger and ringdown phases of coalescence. Second, all the curves are relatively steep, with typical detection SNRs mapping out a small range of $\beta$ for each $\bar{b}$. Indeed, our analytical formulation predicts $\mathrm{SNR}^\mathrm{detect}$ and $\mathrm{SNR}^\mathrm{bias}$ proportional to $\beta^{-1}$ at leading order (this is because $\delta h \propto \beta$, and both $\delta \vec{\theta}_\mathrm{th}$ and $\mathrm{SNR}_\mathrm{MG} \propto \delta h$).
Third, while we computed all our results for positive $\beta$, our formulation is invariant with respect to a change in its sign: with everything else held fixed, the change yields $\delta h \rightarrow -\delta h$ and $\delta \vec{\theta}_\mathrm{th} \rightarrow -\delta \vec{\theta}_\mathrm{th}$, which leaves $\mathrm{SNR}^\mathrm{detect}$ and $\mathrm{SNR}^\mathrm{bias}$ unchanged.
Fourth, and most important, the $\mathrm{SNR}^\mathrm{bias}$ curves lie almost always to the \emph{left} of the $\mathrm{SNR}^\mathrm{detect}$ curves---stealth bias (which corresponds to the area between the solid and dashed line) is a generic feature for these waveforms, and becomes more prevalent for more negative $\bar{b}$. 

Thus, if ground-based GW detectors observe a signal from these binaries with a certain detection SNR, and Nature happens to deviate from GR in the strong-field regime, three cases are possible (with somewhat blurry boundaries) depending on the value of $\beta$. They are illustrated by the small symbols in the top panel of Fig.\ \ref{Main-Figure} and listed below:
\begin{itemize}
\item \textbf{Overt bias: significant fundamental bias, detectable MG deviation}:
$\mathrm{SNR}^\mathrm{bias}(\beta) < \mathrm{SNR}^\mathrm{detect}(\beta) < \mathrm{SNR}$ (rightmost symbol, star). The systematic error induced by using GR templates is larger than the statistical uncertainty, but we are able to infer that the signal is described by a non-GR theory. Thus, we can correct the fundamental bias, as long as our MG model is sufficiently close to the true theory of gravity.
\item \textbf{Stealth bias: significant fundamental bias, undetectable MG deviation}:
$\mathrm{SNR}^\mathrm{bias}(\beta) < \mathrm{SNR} < \mathrm{SNR}^\mathrm{detect}(\beta)$
(middle symbol, square). The systematic error is larger than the statistical uncertainty, but we do not have sufficient statistical evidence to determine that the signal is described by a MG theory. Thus, we cannot correct the fundamental bias. Parameter-estimation results might have to be prefaced with the caveat that they may include a systematic error as large as the statistical uncertainty \emph{if} GR is incorrect in the strong-field with $\beta$ in this range.
\item \textbf{Negligible bias: insignificant fundamental bias, undetectable MG deviation}:
$\mathrm{SNR} < \mathrm{SNR}^\mathrm{bias}(\beta) < \mathrm{SNR}^\mathrm{detect}(\beta)$
(leftmost symbol, circle). Although we cannot determine that the signal is described by a MG theory (i.e., the signal is indistinguishable from the GR prediction), the systematic error is smaller than the statistical uncertainty, so the accuracy of parameter estimation is not affected. This is also trivially the case when GR is correct.
\end{itemize}
By contrast, if we had found that in general $\mathrm{SNR}^\mathrm{bias}(\beta) > \mathrm{SNR}^\mathrm{detect}(\beta)$, systematic error would only appear in observations where the presence of MG corrections is obvious. Although our analysis is robust for the sources we examined, we cannot extend our conclusions to different sources, different waveform families, or different detectors. We leave this to future work.
\setlength{\tabcolsep}{5pt} 
\setlength{\extrarowheight}{1.5pt}
\begin{table*}[ht]
\begin{tabular}{c| c c c c c c c c c}
\hline\hline
$\bar{b}$ & $-7$  & $-6$  & $-5$  & $-4$  & $-3$  & $-2$  & $-1$  & $1$ & $2$\\ 
\hline
upper limit on $\beta$ & $4 \times 10^{-10}$  & $4 \times 10^{-7}$  & $3 \times 10^{-4}$ & $3 \times 10^{-1}$  & $3 \times 10^{2}$  & $3 \times 10^{5}$  & $5 \times 10^{8}$  & $4 \times 10^{14}$ &  $2 \times 10^{17}$\\ 
\hline\hline
\end{tabular}
\caption{\label{prev-cons}Binary pulsar constraints on $\beta$ for different values of $\bar{b}$~\cite{Yunes:2010qb}.}
\end{table*}

Of course, overt and (particularly) stealth bias are only a concern if they are not already excluded \emph{a priori} by previously obtained constraints on $\beta$ for a given $\bar{b}$. For instance, binary-pulsar observations \cite{Yunes:2010qb} provide very strong constraints for $\bar{b} \leq -6$ (see Table \ref{prev-cons}), but deteriorate rapidly for corrections corresponding to higher-order PN terms \cite{2011PhRvD..84f2003C}. For this reason, it would be very interesting to map all current binary-pulsar and Solar-System constraints to the ppE framework. 

\section{Conclusions}
\label{sec:conclusions}

In this paper we introduced a scheme to determine if the systematic errors induced by the \emph{a priori} assumption that GR is always correct could become as significant as noise-induced statistical uncertainties \emph{before} a MG effect could be detected confidently using Bayesian model comparison. We find that indeed such stealth biases appear generically in analysis of inspiral GWs from stellar-mass compact binaries, as observed by second-generation ground-based detectors, at least for certain ranges of detection SNRs and MG correction strengths.

The possibility of stealth bias should \emph{not} be regarded as casting a veil of uncertainty on the interpretation of GW detections. Rather, it is a warning that inferences from GW observations must be considered in the context of all the other evidence for GR as the true theory of gravitation, as well as all other constraints on MG theories. 

For simplicity, we studied non-eccentric, non-precessing systems with no PN amplitude corrections to the waveforms, concentrating only on second-generation ground-based detectors. We employed a simple analytical formulation that is valid for sufficiently large detection SNRs. Obvious extensions of this work could encompass more complete waveform models (perhaps including also the merger and ringdown phases of binary coalescence, using the hybrid ppE models); different source populations; low-SNR detections, which can be tackled with Cutler and Vallisneri's ``ODE'' approach to evaluate fundamental bias \cite{Cutler:2007mi} and with Monte Carlo integration to evaluate the Bayesian odds ratio \cite{2011PhRvD..84f2003C}; space-borne detectors or third-generation ground-based detectors. Indeed, we have no doubt that the full power of these techniques will be unleashed once actual detections are made, to look for hints of MG theories and put GR on an ever-firmer footing. 

\section{Acknowledgements}

We thank Curt Cutler for useful comments and suggestions, Katerina Chatziioannou for providing details about the Picard--Lindel\"{o}f and Cauchy--Lipschitz uniqueness theorems, Neil Cornish and Laura Sampson for help in comparing some of our Fisher calculations with the results of Cornish's MCMC code, and the organizers of the 2011 Astro-GR workshop, where this work was conceived. NY acknowledges support from NSF grant PHY-1114374 and NASA grant NNX11AI49G, under sub-award 00001944. MV's research was supported by NASA grant NNX10AC69G, and was performed at the Jet Propulsion Laboratory under contract with the National Space and Aeronautics Administration. Copyright 2013.
 
\bibliography{phyjabb,master}

\begin{thebibliography}{52}%
\makeatletter
\providecommand \@ifxundefined [1]{%
 \@ifx{#1\undefined}
}%
\providecommand \@ifnum [1]{%
 \ifnum #1\expandafter \@firstoftwo
 \else \expandafter \@secondoftwo
 \fi
}%
\providecommand \@ifx [1]{%
 \ifx #1\expandafter \@firstoftwo
 \else \expandafter \@secondoftwo
 \fi
}%
\providecommand \natexlab [1]{#1}%
\providecommand \enquote  [1]{``#1''}%
\providecommand \bibnamefont  [1]{#1}%
\providecommand \bibfnamefont [1]{#1}%
\providecommand \citenamefont [1]{#1}%
\providecommand \href@noop [0]{\@secondoftwo}%
\providecommand \href [0]{\begingroup \@sanitize@url \@href}%
\providecommand \@href[1]{\@@startlink{#1}\@@href}%
\providecommand \@@href[1]{\endgroup#1\@@endlink}%
\providecommand \@sanitize@url [0]{\catcode `\\12\catcode `\$12\catcode
  `\&12\catcode `\#12\catcode `\^12\catcode `\_12\catcode `\%12\relax}%
\providecommand \@@startlink[1]{}%
\providecommand \@@endlink[0]{}%
\providecommand \url  [0]{\begingroup\@sanitize@url \@url }%
\providecommand \@url [1]{\endgroup\@href {#1}{\urlprefix }}%
\providecommand \urlprefix  [0]{URL }%
\providecommand \Eprint [0]{\href }%
\providecommand \doibase [0]{http://dx.doi.org/}%
\providecommand \selectlanguage [0]{\@gobble}%
\providecommand \bibinfo  [0]{\@secondoftwo}%
\providecommand \bibfield  [0]{\@secondoftwo}%
\providecommand \translation [1]{[#1]}%
\providecommand \BibitemOpen [0]{}%
\providecommand \bibitemStop [0]{}%
\providecommand \bibitemNoStop [0]{.\EOS\space}%
\providecommand \EOS [0]{\spacefactor3000\relax}%
\providecommand \BibitemShut  [1]{\csname bibitem#1\endcsname}%
\let\auto@bib@innerbib\@empty
\bibitem [{\citenamefont {{Will}}(2006)}]{2006LRR.....9....3W}%
  \BibitemOpen
  \bibfield  {author} {\bibinfo {author} {\bibfnamefont {C.~M.}\ \bibnamefont
  {{Will}}},\ }\href@noop {} {\bibfield  {journal} {\bibinfo  {journal} {Living
  Reviews in Relativity}\ }\textbf {\bibinfo {volume} {9}},\ \bibinfo {pages}
  {3} (\bibinfo {year} {2006})},\ \Eprint
  {http://arxiv.org/abs/arXiv:gr-qc/0510072} {arXiv:gr-qc/0510072} \BibitemShut
  {NoStop}%
\bibitem [{\citenamefont {{Yunes}}\ and\ \citenamefont
  {{Pretorius}}(2009{\natexlab{a}})}]{2009PhRvD..80l2003Y}%
  \BibitemOpen
  \bibfield  {author} {\bibinfo {author} {\bibfnamefont {N.}~\bibnamefont
  {{Yunes}}}\ and\ \bibinfo {author} {\bibfnamefont {F.}~\bibnamefont
  {{Pretorius}}},\ }\href {\doibase 10.1103/PhysRevD.80.122003} {\bibfield
  {journal} {\bibinfo  {journal} {\prd}\ }\textbf {\bibinfo {volume} {80}},\
  \bibinfo {pages} {122003} (\bibinfo {year} {2009}{\natexlab{a}})},\ \Eprint
  {http://arxiv.org/abs/0909.3328} {arXiv:0909.3328 [gr-qc]} \BibitemShut
  {NoStop}%
\bibitem [{\citenamefont {{Cornish}}\ \emph {et~al.}(2011)\citenamefont
  {{Cornish}}, \citenamefont {{Sampson}}, \citenamefont {{Yunes}},\ and\
  \citenamefont {{Pretorius}}}]{2011PhRvD..84f2003C}%
  \BibitemOpen
  \bibfield  {author} {\bibinfo {author} {\bibfnamefont {N.}~\bibnamefont
  {{Cornish}}}, \bibinfo {author} {\bibfnamefont {L.}~\bibnamefont
  {{Sampson}}}, \bibinfo {author} {\bibfnamefont {N.}~\bibnamefont {{Yunes}}},
  \ and\ \bibinfo {author} {\bibfnamefont {F.}~\bibnamefont {{Pretorius}}},\
  }\href {\doibase 10.1103/PhysRevD.84.062003} {\bibfield  {journal} {\bibinfo
  {journal} {\prd}\ }\textbf {\bibinfo {volume} {84}},\ \bibinfo {pages}
  {062003} (\bibinfo {year} {2011})},\ \Eprint {http://arxiv.org/abs/1105.2088}
  {arXiv:1105.2088 [gr-qc]} \BibitemShut {NoStop}%
\bibitem [{\citenamefont {{Li}}\ \emph {et~al.}(2012)\citenamefont {{Li}},
  \citenamefont {{Del Pozzo}}, \citenamefont {{Vitale}}, \citenamefont {{Van
  Den Broeck}}, \citenamefont {{Agathos}}, \citenamefont {{Veitch}},
  \citenamefont {{Grover}}, \citenamefont {{Sidery}}, \citenamefont
  {{Sturani}},\ and\ \citenamefont {{Vecchio}}}]{2012PhRvD..85h2003L}%
  \BibitemOpen
  \bibfield  {author} {\bibinfo {author} {\bibfnamefont {T.~G.~F.}\
  \bibnamefont {{Li}}}, \bibinfo {author} {\bibfnamefont {W.}~\bibnamefont
  {{Del Pozzo}}}, \bibinfo {author} {\bibfnamefont {S.}~\bibnamefont
  {{Vitale}}}, \bibinfo {author} {\bibfnamefont {C.}~\bibnamefont {{Van Den
  Broeck}}}, \bibinfo {author} {\bibfnamefont {M.}~\bibnamefont {{Agathos}}},
  \bibinfo {author} {\bibfnamefont {J.}~\bibnamefont {{Veitch}}}, \bibinfo
  {author} {\bibfnamefont {K.}~\bibnamefont {{Grover}}}, \bibinfo {author}
  {\bibfnamefont {T.}~\bibnamefont {{Sidery}}}, \bibinfo {author}
  {\bibfnamefont {R.}~\bibnamefont {{Sturani}}}, \ and\ \bibinfo {author}
  {\bibfnamefont {A.}~\bibnamefont {{Vecchio}}},\ }\href {\doibase
  10.1103/PhysRevD.85.082003} {\bibfield  {journal} {\bibinfo  {journal}
  {\prd}\ }\textbf {\bibinfo {volume} {85}},\ \bibinfo {eid} {082003} (\bibinfo
  {year} {2012})},\ \Eprint {http://arxiv.org/abs/1110.0530} {arXiv:1110.0530
  [gr-qc]} \BibitemShut {NoStop}%
\bibitem [{\citenamefont {Arun}\ \emph
  {et~al.}(2006{\natexlab{a}})\citenamefont {Arun}, \citenamefont {Iyer},
  \citenamefont {Qusailah},\ and\ \citenamefont {Sathyaprakash}}]{Arun:2006hn}%
  \BibitemOpen
  \bibfield  {author} {\bibinfo {author} {\bibfnamefont {K.~G.}\ \bibnamefont
  {Arun}}, \bibinfo {author} {\bibfnamefont {B.~R.}\ \bibnamefont {Iyer}},
  \bibinfo {author} {\bibfnamefont {M.~S.~S.}\ \bibnamefont {Qusailah}}, \ and\
  \bibinfo {author} {\bibfnamefont {B.~S.}\ \bibnamefont {Sathyaprakash}},\
  }\href {\doibase 10.1103/PhysRevD.74.024006} {\bibfield  {journal} {\bibinfo
  {journal} {Phys. Rev.}\ }\textbf {\bibinfo {volume} {D74}},\ \bibinfo {pages}
  {024006} (\bibinfo {year} {2006}{\natexlab{a}})},\ \Eprint
  {http://arxiv.org/abs/gr-qc/0604067} {arXiv:gr-qc/0604067} \BibitemShut
  {NoStop}%
\bibitem [{\citenamefont {Arun}\ \emph
  {et~al.}(2006{\natexlab{b}})\citenamefont {Arun}, \citenamefont {Iyer},
  \citenamefont {Qusailah},\ and\ \citenamefont {Sathyaprakash}}]{Arun:2006yw}%
  \BibitemOpen
  \bibfield  {author} {\bibinfo {author} {\bibfnamefont {K.~G.}\ \bibnamefont
  {Arun}}, \bibinfo {author} {\bibfnamefont {B.~R.}\ \bibnamefont {Iyer}},
  \bibinfo {author} {\bibfnamefont {M.~S.~S.}\ \bibnamefont {Qusailah}}, \ and\
  \bibinfo {author} {\bibfnamefont {B.~S.}\ \bibnamefont {Sathyaprakash}},\
  }\href@noop {} {\bibfield  {journal} {\bibinfo  {journal} {Class. Quantum
  Grav.}\ }\textbf {\bibinfo {volume} {23}},\ \bibinfo {pages} {L37} (\bibinfo
  {year} {2006}{\natexlab{b}})},\ \Eprint {http://arxiv.org/abs/gr-qc/0604018}
  {arXiv:gr-qc/0604018} \BibitemShut {NoStop}%
\bibitem [{\citenamefont {{Mishra}}\ \emph {et~al.}(2010)\citenamefont
  {{Mishra}}, \citenamefont {{Arun}}, \citenamefont {{Iyer}},\ and\
  \citenamefont {{Sathyaprakash}}}]{2010PhRvD..82f4010M}%
  \BibitemOpen
  \bibfield  {author} {\bibinfo {author} {\bibfnamefont {C.~K.}\ \bibnamefont
  {{Mishra}}}, \bibinfo {author} {\bibfnamefont {K.~G.}\ \bibnamefont
  {{Arun}}}, \bibinfo {author} {\bibfnamefont {B.~R.}\ \bibnamefont {{Iyer}}},
  \ and\ \bibinfo {author} {\bibfnamefont {B.~S.}\ \bibnamefont
  {{Sathyaprakash}}},\ }\href {\doibase 10.1103/PhysRevD.82.064010} {\bibfield
  {journal} {\bibinfo  {journal} {\prd}\ }\textbf {\bibinfo {volume} {82}},\
  \bibinfo {pages} {064010} (\bibinfo {year} {2010})},\ \Eprint
  {http://arxiv.org/abs/1005.0304} {arXiv:1005.0304 [gr-qc]} \BibitemShut
  {NoStop}%
\bibitem [{\citenamefont {Will}(1994)}]{Will:1994fb}%
  \BibitemOpen
  \bibfield  {author} {\bibinfo {author} {\bibfnamefont {C.~M.}\ \bibnamefont
  {Will}},\ }\href {\doibase 10.1103/PhysRevD.50.6058} {\bibfield  {journal}
  {\bibinfo  {journal} {Phys. Rev. D}\ }\textbf {\bibinfo {volume} {50}},\
  \bibinfo {pages} {6058} (\bibinfo {year} {1994})},\ \Eprint
  {http://arxiv.org/abs/gr-qc/9406022} {arXiv:gr-qc/9406022} \BibitemShut
  {NoStop}%
\bibitem [{\citenamefont {Will}\ and\ \citenamefont
  {Yunes}(2004)}]{Will:2004xi}%
  \BibitemOpen
  \bibfield  {author} {\bibinfo {author} {\bibfnamefont {C.~M.}\ \bibnamefont
  {Will}}\ and\ \bibinfo {author} {\bibfnamefont {N.}~\bibnamefont {Yunes}},\
  }\href {\doibase 10.1088/0264-9381/21/18/006} {\bibfield  {journal} {\bibinfo
   {journal} {Class. Quantum Grav.}\ }\textbf {\bibinfo {volume} {21}},\
  \bibinfo {pages} {4367} (\bibinfo {year} {2004})},\ \Eprint
  {http://arxiv.org/abs/gr-qc/0403100} {arXiv:gr-qc/0403100} \BibitemShut
  {NoStop}%
\bibitem [{\citenamefont {Berti}\ \emph
  {et~al.}(2005{\natexlab{a}})\citenamefont {Berti}, \citenamefont {Buonanno},\
  and\ \citenamefont {Will}}]{Berti:2005qd}%
  \BibitemOpen
  \bibfield  {author} {\bibinfo {author} {\bibfnamefont {E.}~\bibnamefont
  {Berti}}, \bibinfo {author} {\bibfnamefont {A.}~\bibnamefont {Buonanno}}, \
  and\ \bibinfo {author} {\bibfnamefont {C.~M.}\ \bibnamefont {Will}},\ }\href
  {\doibase 10.1088/0264-9381/22/18/S08} {\bibfield  {journal} {\bibinfo
  {journal} {Class. Quantum Grav.}\ }\textbf {\bibinfo {volume} {22}},\
  \bibinfo {pages} {S943} (\bibinfo {year} {2005}{\natexlab{a}})},\ \Eprint
  {http://arxiv.org/abs/gr-qc/0504017} {arXiv:gr-qc/0504017} \BibitemShut
  {NoStop}%
\bibitem [{\citenamefont {{Stavridis}}\ and\ \citenamefont
  {{Will}}(2009)}]{2009PhRvD..80d4002S}%
  \BibitemOpen
  \bibfield  {author} {\bibinfo {author} {\bibfnamefont {A.}~\bibnamefont
  {{Stavridis}}}\ and\ \bibinfo {author} {\bibfnamefont {C.~M.}\ \bibnamefont
  {{Will}}},\ }\href {\doibase 10.1103/PhysRevD.80.044002} {\bibfield
  {journal} {\bibinfo  {journal} {\prd}\ }\textbf {\bibinfo {volume} {80}},\
  \bibinfo {eid} {044002} (\bibinfo {year} {2009})},\ \Eprint
  {http://arxiv.org/abs/0906.3602} {arXiv:0906.3602 [gr-qc]} \BibitemShut
  {NoStop}%
\bibitem [{\citenamefont {{Arun}}\ and\ \citenamefont
  {{Will}}(2009)}]{2009CQGra..26o5002A}%
  \BibitemOpen
  \bibfield  {author} {\bibinfo {author} {\bibfnamefont {K.~G.}\ \bibnamefont
  {{Arun}}}\ and\ \bibinfo {author} {\bibfnamefont {C.~M.}\ \bibnamefont
  {{Will}}},\ }\href {\doibase 10.1088/0264-9381/26/15/155002} {\bibfield
  {journal} {\bibinfo  {journal} {Class. Quantum Grav.}\ }\textbf {\bibinfo
  {volume} {26}},\ \bibinfo {pages} {155002} (\bibinfo {year} {2009})},\
  \Eprint {http://arxiv.org/abs/0904.1190} {arXiv:0904.1190 [gr-qc]}
  \BibitemShut {NoStop}%
\bibitem [{\citenamefont {Keppel}\ and\ \citenamefont
  {Ajith}(2010)}]{Keppel:2010qu}%
  \BibitemOpen
  \bibfield  {author} {\bibinfo {author} {\bibfnamefont {D.}~\bibnamefont
  {Keppel}}\ and\ \bibinfo {author} {\bibfnamefont {P.}~\bibnamefont {Ajith}},\
  }\href {\doibase 10.1103/PhysRevD.82.122001} {\bibfield  {journal} {\bibinfo
  {journal} {Phys. Rev. D}\ }\textbf {\bibinfo {volume} {82}},\ \bibinfo
  {pages} {122001} (\bibinfo {year} {2010})},\ \Eprint
  {http://arxiv.org/abs/1004.0284} {arXiv:1004.0284 [gr-qc]} \BibitemShut
  {NoStop}%
\bibitem [{\citenamefont {{Yagi}}\ and\ \citenamefont
  {{Tanaka}}(2010)}]{2010PhRvD..81f4008Y}%
  \BibitemOpen
  \bibfield  {author} {\bibinfo {author} {\bibfnamefont {K.}~\bibnamefont
  {{Yagi}}}\ and\ \bibinfo {author} {\bibfnamefont {T.}~\bibnamefont
  {{Tanaka}}},\ }\href {\doibase 10.1103/PhysRevD.81.064008} {\bibfield
  {journal} {\bibinfo  {journal} {\prd}\ }\textbf {\bibinfo {volume} {81}},\
  \bibinfo {eid} {064008} (\bibinfo {year} {2010})},\ \Eprint
  {http://arxiv.org/abs/0906.4269} {arXiv:0906.4269 [gr-qc]} \BibitemShut
  {NoStop}%
\bibitem [{\citenamefont {Yunes}\ \emph {et~al.}(2012)\citenamefont {Yunes},
  \citenamefont {Pani},\ and\ \citenamefont {Cardoso}}]{Yunes:2011aa}%
  \BibitemOpen
  \bibfield  {author} {\bibinfo {author} {\bibfnamefont {N.}~\bibnamefont
  {Yunes}}, \bibinfo {author} {\bibfnamefont {P.}~\bibnamefont {Pani}}, \ and\
  \bibinfo {author} {\bibfnamefont {V.}~\bibnamefont {Cardoso}},\ }\href
  {\doibase 10.1103/PhysRevD.85.102003} {\bibfield  {journal} {\bibinfo
  {journal} {Phys. Rev. D}\ }\textbf {\bibinfo {volume} {85}},\ \bibinfo
  {pages} {102003} (\bibinfo {year} {2012})},\ \Eprint
  {http://arxiv.org/abs/1112.3351} {arXiv:1112.3351 [gr-qc]} \BibitemShut
  {NoStop}%
\bibitem [{\citenamefont {Healy}\ \emph {et~al.}(2011)\citenamefont {Healy},
  \citenamefont {Bode}, \citenamefont {Haas}, \citenamefont {Pazos},
  \citenamefont {Laguna} \emph {et~al.}}]{Healy:2011ef}%
  \BibitemOpen
  \bibfield  {author} {\bibinfo {author} {\bibfnamefont {J.}~\bibnamefont
  {Healy}}, \bibinfo {author} {\bibfnamefont {T.}~\bibnamefont {Bode}},
  \bibinfo {author} {\bibfnamefont {R.}~\bibnamefont {Haas}}, \bibinfo {author}
  {\bibfnamefont {E.}~\bibnamefont {Pazos}}, \bibinfo {author} {\bibfnamefont
  {P.}~\bibnamefont {Laguna}},  \emph {et~al.},\ }\href@noop {} {\  (\bibinfo
  {year} {2011})},\ \Eprint {http://arxiv.org/abs/1112.3928} {arXiv:1112.3928
  [gr-qc]} \BibitemShut {NoStop}%
\bibitem [{\citenamefont {{Sopuerta}}\ and\ \citenamefont
  {{Yunes}}(2009)}]{Sopuerta:2009iy}%
  \BibitemOpen
  \bibfield  {author} {\bibinfo {author} {\bibfnamefont {C.~F.}\ \bibnamefont
  {{Sopuerta}}}\ and\ \bibinfo {author} {\bibfnamefont {N.}~\bibnamefont
  {{Yunes}}},\ }\href {\doibase 10.1103/PhysRevD.80.064006} {\bibfield
  {journal} {\bibinfo  {journal} {\prd}\ }\textbf {\bibinfo {volume} {80}},\
  \bibinfo {eid} {064006} (\bibinfo {year} {2009})},\ \Eprint
  {http://arxiv.org/abs/0904.4501} {arXiv:0904.4501 [gr-qc]} \BibitemShut
  {NoStop}%
\bibitem [{\citenamefont {{Yunes}}\ and\ \citenamefont
  {{Pretorius}}(2009{\natexlab{b}})}]{Yunes:2009hc}%
  \BibitemOpen
  \bibfield  {author} {\bibinfo {author} {\bibfnamefont {N.}~\bibnamefont
  {{Yunes}}}\ and\ \bibinfo {author} {\bibfnamefont {F.}~\bibnamefont
  {{Pretorius}}},\ }\href {\doibase 10.1103/PhysRevD.79.084043} {\bibfield
  {journal} {\bibinfo  {journal} {\prd}\ }\textbf {\bibinfo {volume} {79}},\
  \bibinfo {eid} {084043} (\bibinfo {year} {2009}{\natexlab{b}})},\ \Eprint
  {http://arxiv.org/abs/0902.4669} {arXiv:0902.4669 [gr-qc]} \BibitemShut
  {NoStop}%
\bibitem [{\citenamefont {{Yagi}}\ \emph {et~al.}(2012)\citenamefont {{Yagi}},
  \citenamefont {{Stein}}, \citenamefont {{Yunes}},\ and\ \citenamefont
  {{Tanaka}}}]{2012PhRvD..85f4022Y}%
  \BibitemOpen
  \bibfield  {author} {\bibinfo {author} {\bibfnamefont {K.}~\bibnamefont
  {{Yagi}}}, \bibinfo {author} {\bibfnamefont {L.~C.}\ \bibnamefont {{Stein}}},
  \bibinfo {author} {\bibfnamefont {N.}~\bibnamefont {{Yunes}}}, \ and\
  \bibinfo {author} {\bibfnamefont {T.}~\bibnamefont {{Tanaka}}},\ }\href
  {\doibase 10.1103/PhysRevD.85.064022} {\bibfield  {journal} {\bibinfo
  {journal} {\prd}\ }\textbf {\bibinfo {volume} {85}},\ \bibinfo {eid} {064022}
  (\bibinfo {year} {2012})},\ \Eprint {http://arxiv.org/abs/1110.5950}
  {arXiv:1110.5950 [gr-qc]} \BibitemShut {NoStop}%
\bibitem [{\citenamefont {Yagi}\ \emph
  {et~al.}(2012{\natexlab{a}})\citenamefont {Yagi}, \citenamefont {Yunes},\
  and\ \citenamefont {Tanaka}}]{Yagi:2012ya}%
  \BibitemOpen
  \bibfield  {author} {\bibinfo {author} {\bibfnamefont {K.}~\bibnamefont
  {Yagi}}, \bibinfo {author} {\bibfnamefont {N.}~\bibnamefont {Yunes}}, \ and\
  \bibinfo {author} {\bibfnamefont {T.}~\bibnamefont {Tanaka}},\ }\href
  {\doibase 10.1103/PhysRevD.86.044037} {\bibfield  {journal} {\bibinfo
  {journal} {Phys. Rev. D}\ }\textbf {\bibinfo {volume} {86}},\ \bibinfo
  {pages} {044037} (\bibinfo {year} {2012}{\natexlab{a}})},\ \Eprint
  {http://arxiv.org/abs/1206.6130} {arXiv:1206.6130 [gr-qc]} \BibitemShut
  {NoStop}%
\bibitem [{\citenamefont {Yagi}\ \emph
  {et~al.}(2012{\natexlab{b}})\citenamefont {Yagi}, \citenamefont {Yunes},\
  and\ \citenamefont {Tanaka}}]{Yagi:2012vf}%
  \BibitemOpen
  \bibfield  {author} {\bibinfo {author} {\bibfnamefont {K.}~\bibnamefont
  {Yagi}}, \bibinfo {author} {\bibfnamefont {N.}~\bibnamefont {Yunes}}, \ and\
  \bibinfo {author} {\bibfnamefont {T.}~\bibnamefont {Tanaka}},\ }\href@noop {}
  {\  (\bibinfo {year} {2012}{\natexlab{b}})},\ \Eprint
  {http://arxiv.org/abs/1208.5102} {arXiv:1208.5102 [gr-qc]} \BibitemShut
  {NoStop}%
\bibitem [{\citenamefont {Will}(1998)}]{Will:1997bb}%
  \BibitemOpen
  \bibfield  {author} {\bibinfo {author} {\bibfnamefont {C.~M.}\ \bibnamefont
  {Will}},\ }\href {\doibase 10.1103/PhysRevD.57.2061} {\bibfield  {journal}
  {\bibinfo  {journal} {Phys. Rev. D}\ }\textbf {\bibinfo {volume} {57}},\
  \bibinfo {pages} {2061} (\bibinfo {year} {1998})},\ \Eprint
  {http://arxiv.org/abs/gr-qc/9709011} {arXiv:gr-qc/9709011} \BibitemShut
  {NoStop}%
\bibitem [{\citenamefont {Scharre}\ and\ \citenamefont
  {Will}(2002)}]{Scharre:2001hn}%
  \BibitemOpen
  \bibfield  {author} {\bibinfo {author} {\bibfnamefont {P.~D.}\ \bibnamefont
  {Scharre}}\ and\ \bibinfo {author} {\bibfnamefont {C.~M.}\ \bibnamefont
  {Will}},\ }\href {\doibase 10.1103/PhysRevD.65.042002} {\bibfield  {journal}
  {\bibinfo  {journal} {Phys. Rev. D}\ }\textbf {\bibinfo {volume} {65}},\
  \bibinfo {pages} {042002} (\bibinfo {year} {2002})},\ \Eprint
  {http://arxiv.org/abs/gr-qc/0109044} {arXiv:gr-qc/0109044} \BibitemShut
  {NoStop}%
\bibitem [{\citenamefont {{Berti}}\ \emph {et~al.}(2011)\citenamefont
  {{Berti}}, \citenamefont {{Gair}},\ and\ \citenamefont
  {{Sesana}}}]{2011PhRvD..84j1501B}%
  \BibitemOpen
  \bibfield  {author} {\bibinfo {author} {\bibfnamefont {E.}~\bibnamefont
  {{Berti}}}, \bibinfo {author} {\bibfnamefont {J.}~\bibnamefont {{Gair}}}, \
  and\ \bibinfo {author} {\bibfnamefont {A.}~\bibnamefont {{Sesana}}},\ }\href
  {\doibase 10.1103/PhysRevD.84.101501} {\bibfield  {journal} {\bibinfo
  {journal} {\prd}\ }\textbf {\bibinfo {volume} {84}},\ \bibinfo {eid} {101501}
  (\bibinfo {year} {2011})},\ \Eprint {http://arxiv.org/abs/1107.3528}
  {arXiv:1107.3528 [gr-qc]} \BibitemShut {NoStop}%
\bibitem [{\citenamefont {{Mirshekari}}\ \emph {et~al.}(2012)\citenamefont
  {{Mirshekari}}, \citenamefont {{Yunes}},\ and\ \citenamefont
  {{Will}}}]{2012PhRvD..85b4041M}%
  \BibitemOpen
  \bibfield  {author} {\bibinfo {author} {\bibfnamefont {S.}~\bibnamefont
  {{Mirshekari}}}, \bibinfo {author} {\bibfnamefont {N.}~\bibnamefont
  {{Yunes}}}, \ and\ \bibinfo {author} {\bibfnamefont {C.~M.}\ \bibnamefont
  {{Will}}},\ }\href {\doibase 10.1103/PhysRevD.85.024041} {\bibfield
  {journal} {\bibinfo  {journal} {\prd}\ }\textbf {\bibinfo {volume} {85}},\
  \bibinfo {eid} {024041} (\bibinfo {year} {2012})},\ \Eprint
  {http://arxiv.org/abs/1110.2720} {arXiv:1110.2720 [gr-qc]} \BibitemShut
  {NoStop}%
\bibitem [{\citenamefont {Yunes}\ \emph {et~al.}(2010)\citenamefont {Yunes},
  \citenamefont {O'Shaughnessy}, \citenamefont {Owen},\ and\ \citenamefont
  {Alexander}}]{Yunes:2010yf}%
  \BibitemOpen
  \bibfield  {author} {\bibinfo {author} {\bibfnamefont {N.}~\bibnamefont
  {Yunes}}, \bibinfo {author} {\bibfnamefont {R.}~\bibnamefont
  {O'Shaughnessy}}, \bibinfo {author} {\bibfnamefont {B.~J.}\ \bibnamefont
  {Owen}}, \ and\ \bibinfo {author} {\bibfnamefont {S.}~\bibnamefont
  {Alexander}},\ }\href {\doibase 10.1103/PhysRevD.82.064017} {\bibfield
  {journal} {\bibinfo  {journal} {Phys. Rev. D}\ }\textbf {\bibinfo {volume}
  {82}},\ \bibinfo {pages} {064017} (\bibinfo {year} {2010})},\ \Eprint
  {http://arxiv.org/abs/1005.3310} {arXiv:1005.3310 [gr-qc]} \BibitemShut
  {NoStop}%
\bibitem [{\citenamefont {{Yunes}}\ \emph {et~al.}(2010)\citenamefont
  {{Yunes}}, \citenamefont {{Pretorius}},\ and\ \citenamefont
  {{Spergel}}}]{Yunes:2009bv}%
  \BibitemOpen
  \bibfield  {author} {\bibinfo {author} {\bibfnamefont {N.}~\bibnamefont
  {{Yunes}}}, \bibinfo {author} {\bibfnamefont {F.}~\bibnamefont
  {{Pretorius}}}, \ and\ \bibinfo {author} {\bibfnamefont {D.}~\bibnamefont
  {{Spergel}}},\ }\href {\doibase 10.1103/PhysRevD.81.064018} {\bibfield
  {journal} {\bibinfo  {journal} {\prd}\ }\textbf {\bibinfo {volume} {81}},\
  \bibinfo {pages} {064018} (\bibinfo {year} {2010})},\ \Eprint
  {http://arxiv.org/abs/0912.2724} {arXiv:0912.2724 [gr-qc]} \BibitemShut
  {NoStop}%
\bibitem [{\citenamefont {{Yagi}}\ \emph {et~al.}(2011)\citenamefont {{Yagi}},
  \citenamefont {{Tanahashi}},\ and\ \citenamefont {{Tanaka}}}]{Yagi:2011yu}%
  \BibitemOpen
  \bibfield  {author} {\bibinfo {author} {\bibfnamefont {K.}~\bibnamefont
  {{Yagi}}}, \bibinfo {author} {\bibfnamefont {N.}~\bibnamefont {{Tanahashi}}},
  \ and\ \bibinfo {author} {\bibfnamefont {T.}~\bibnamefont {{Tanaka}}},\
  }\href {\doibase 10.1103/PhysRevD.83.084036} {\bibfield  {journal} {\bibinfo
  {journal} {\prd}\ }\textbf {\bibinfo {volume} {83}},\ \bibinfo {pages}
  {084036} (\bibinfo {year} {2011})},\ \Eprint {http://arxiv.org/abs/1101.4997}
  {arXiv:1101.4997 [gr-qc]} \BibitemShut {NoStop}%
\bibitem [{\citenamefont {{Gregory}}(2005)}]{2005blda.book.....G}%
  \BibitemOpen
  \bibfield  {author} {\bibinfo {author} {\bibfnamefont {P.~C.}\ \bibnamefont
  {{Gregory}}},\ }\href@noop {} {\emph {\bibinfo {title} {{Bayesian Logical
  Data Analysis for the Physical Sciences}}}}\ (\bibinfo  {publisher}
  {Cambridge University Press},\ \bibinfo {year} {2005})\BibitemShut {NoStop}%
\bibitem [{\citenamefont {Cutler}\ and\ \citenamefont
  {Vallisneri}(2007)}]{Cutler:2007mi}%
  \BibitemOpen
  \bibfield  {author} {\bibinfo {author} {\bibfnamefont {C.}~\bibnamefont
  {Cutler}}\ and\ \bibinfo {author} {\bibfnamefont {M.}~\bibnamefont
  {Vallisneri}},\ }\href {\doibase 10.1103/PhysRevD.76.104018} {\bibfield
  {journal} {\bibinfo  {journal} {Phys. Rev.}\ }\textbf {\bibinfo {volume}
  {D76}},\ \bibinfo {pages} {104018} (\bibinfo {year} {2007})},\ \Eprint
  {http://arxiv.org/abs/0707.2982} {arXiv:0707.2982 [gr-qc]} \BibitemShut
  {NoStop}%
\bibitem [{\citenamefont {{Vallisneri}}(2012)}]{2012PhRvD..86h2001V}%
  \BibitemOpen
  \bibfield  {author} {\bibinfo {author} {\bibfnamefont {M.}~\bibnamefont
  {{Vallisneri}}},\ }\href {\doibase 10.1103/PhysRevD.86.082001} {\bibfield
  {journal} {\bibinfo  {journal} {\prd}\ }\textbf {\bibinfo {volume} {86}},\
  \bibinfo {eid} {082001} (\bibinfo {year} {2012})},\ \Eprint
  {http://arxiv.org/abs/1207.4759} {arXiv:1207.4759 [gr-qc]} \BibitemShut
  {NoStop}%
\bibitem [{\citenamefont {Misner}\ \emph {et~al.}(1973)\citenamefont {Misner},
  \citenamefont {Thorne},\ and\ \citenamefont {Wheeler}}]{Misner:1973cw}%
  \BibitemOpen
  \bibfield  {author} {\bibinfo {author} {\bibfnamefont {C.~W.}\ \bibnamefont
  {Misner}}, \bibinfo {author} {\bibfnamefont {K.}~\bibnamefont {Thorne}}, \
  and\ \bibinfo {author} {\bibfnamefont {J.~A.}\ \bibnamefont {Wheeler}},\
  }\href@noop {} {\emph {\bibinfo {title} {Gravitation}}}\ (\bibinfo
  {publisher} {W. H. Freeman \& Co.},\ \bibinfo {address} {San Francisco},\
  \bibinfo {year} {1973})\BibitemShut {NoStop}%
\bibitem [{\citenamefont {{Yunes}}\ \emph {et~al.}(2011)\citenamefont
  {{Yunes}}, \citenamefont {{Kocsis}}, \citenamefont {{Loeb}},\ and\
  \citenamefont {{Haiman}}}]{Yunes:2011ws}%
  \BibitemOpen
  \bibfield  {author} {\bibinfo {author} {\bibfnamefont {N.}~\bibnamefont
  {{Yunes}}}, \bibinfo {author} {\bibfnamefont {B.}~\bibnamefont {{Kocsis}}},
  \bibinfo {author} {\bibfnamefont {A.}~\bibnamefont {{Loeb}}}, \ and\ \bibinfo
  {author} {\bibfnamefont {Z.}~\bibnamefont {{Haiman}}},\ }\href {\doibase
  10.1103/PhysRevLett.107.171103} {\bibfield  {journal} {\bibinfo  {journal}
  {Phys. Rev. Lett.}\ }\textbf {\bibinfo {volume} {107}},\ \bibinfo {eid}
  {171103} (\bibinfo {year} {2011})},\ \Eprint {http://arxiv.org/abs/1103.4609}
  {arXiv:1103.4609 [astro-ph.CO]} \BibitemShut {NoStop}%
\bibitem [{\citenamefont {{Vitale}}\ \emph {et~al.}(2012)\citenamefont
  {{Vitale}}, \citenamefont {{Del Pozzo}}, \citenamefont {{Li}}, \citenamefont
  {{Van Den Broeck}}, \citenamefont {{Mandel}}, \citenamefont {{Aylott}},\ and\
  \citenamefont {{Veitch}}}]{Vitale:2011wu}%
  \BibitemOpen
  \bibfield  {author} {\bibinfo {author} {\bibfnamefont {S.}~\bibnamefont
  {{Vitale}}}, \bibinfo {author} {\bibfnamefont {W.}~\bibnamefont {{Del
  Pozzo}}}, \bibinfo {author} {\bibfnamefont {T.~G.~F.}\ \bibnamefont {{Li}}},
  \bibinfo {author} {\bibfnamefont {C.}~\bibnamefont {{Van Den Broeck}}},
  \bibinfo {author} {\bibfnamefont {I.}~\bibnamefont {{Mandel}}}, \bibinfo
  {author} {\bibfnamefont {B.}~\bibnamefont {{Aylott}}}, \ and\ \bibinfo
  {author} {\bibfnamefont {J.}~\bibnamefont {{Veitch}}},\ }\href {\doibase
  10.1103/PhysRevD.85.064034} {\bibfield  {journal} {\bibinfo  {journal}
  {\prd}\ }\textbf {\bibinfo {volume} {85}},\ \bibinfo {eid} {064034} (\bibinfo
  {year} {2012})},\ \Eprint {http://arxiv.org/abs/1111.3044} {arXiv:1111.3044
  [gr-qc]} \BibitemShut {NoStop}%
\bibitem [{\citenamefont {Whelan}\ \emph {et~al.}(2012)\citenamefont {Whelan},
  \citenamefont {Robinson}, \citenamefont {Romano},\ and\ \citenamefont
  {Thrane}}]{Whelan:2012ur}%
  \BibitemOpen
  \bibfield  {author} {\bibinfo {author} {\bibfnamefont {J.~T.}\ \bibnamefont
  {Whelan}}, \bibinfo {author} {\bibfnamefont {E.~L.}\ \bibnamefont
  {Robinson}}, \bibinfo {author} {\bibfnamefont {J.~D.}\ \bibnamefont
  {Romano}}, \ and\ \bibinfo {author} {\bibfnamefont {E.~H.}\ \bibnamefont
  {Thrane}},\ }\href@noop {} {\  (\bibinfo {year} {2012})},\ \Eprint
  {http://arxiv.org/abs/1205.3112} {arXiv:1205.3112 [gr-qc]} \BibitemShut
  {NoStop}%
\bibitem [{\citenamefont {Yunes}\ \emph {et~al.}(2011)\citenamefont {Yunes},
  \citenamefont {Coleman~Miller},\ and\ \citenamefont
  {Thornburg}}]{Yunes:2010sm}%
  \BibitemOpen
  \bibfield  {author} {\bibinfo {author} {\bibfnamefont {N.}~\bibnamefont
  {Yunes}}, \bibinfo {author} {\bibfnamefont {M.}~\bibnamefont
  {Coleman~Miller}}, \ and\ \bibinfo {author} {\bibfnamefont {J.}~\bibnamefont
  {Thornburg}},\ }\href {\doibase 10.1103/PhysRevD.83.044030} {\bibfield
  {journal} {\bibinfo  {journal} {Phys. Rev. D}\ }\textbf {\bibinfo {volume}
  {83}},\ \bibinfo {pages} {044030} (\bibinfo {year} {2011})},\ \Eprint
  {http://arxiv.org/abs/1010.1721} {arXiv:1010.1721 [astro-ph.GA]} \BibitemShut
  {NoStop}%
\bibitem [{\citenamefont {Kocsis}\ \emph {et~al.}(2011)\citenamefont {Kocsis},
  \citenamefont {Yunes},\ and\ \citenamefont {Loeb}}]{Kocsis:2011dr}%
  \BibitemOpen
  \bibfield  {author} {\bibinfo {author} {\bibfnamefont {B.}~\bibnamefont
  {Kocsis}}, \bibinfo {author} {\bibfnamefont {N.}~\bibnamefont {Yunes}}, \
  and\ \bibinfo {author} {\bibfnamefont {A.}~\bibnamefont {Loeb}},\ }\href
  {\doibase 10.1103/PhysRevD.86.049907, 10.1103/PhysRevD.84.024032} {\bibfield
  {journal} {\bibinfo  {journal} {Phys. Rev. D}\ }\textbf {\bibinfo {volume}
  {84}},\ \bibinfo {pages} {024032} (\bibinfo {year} {2011})},\ \Eprint
  {http://arxiv.org/abs/1104.2322} {arXiv:1104.2322 [astro-ph.GA]} \BibitemShut
  {NoStop}%
\bibitem [{\citenamefont {{Cutler}}\ and\ \citenamefont
  {{Flanagan}}(1994)}]{1994PhRvD..49.2658C}%
  \BibitemOpen
  \bibfield  {author} {\bibinfo {author} {\bibfnamefont {C.}~\bibnamefont
  {{Cutler}}}\ and\ \bibinfo {author} {\bibfnamefont {{\'E}.~E.}\ \bibnamefont
  {{Flanagan}}},\ }\href {\doibase 10.1103/PhysRevD.49.2658} {\bibfield
  {journal} {\bibinfo  {journal} {\prd}\ }\textbf {\bibinfo {volume} {49}},\
  \bibinfo {pages} {2658} (\bibinfo {year} {1994})},\ \Eprint
  {http://arxiv.org/abs/arXiv:gr-qc/9402014} {arXiv:gr-qc/9402014} \BibitemShut
  {NoStop}%
\bibitem [{\citenamefont {Flanagan}\ and\ \citenamefont
  {Hughes}(1998{\natexlab{a}})}]{Flanagan:1997sx}%
  \BibitemOpen
  \bibfield  {author} {\bibinfo {author} {\bibfnamefont {E.~E.}\ \bibnamefont
  {Flanagan}}\ and\ \bibinfo {author} {\bibfnamefont {S.~A.}\ \bibnamefont
  {Hughes}},\ }\href {\doibase 10.1103/PhysRevD.57.4535} {\bibfield  {journal}
  {\bibinfo  {journal} {Phys. Rev. D}\ }\textbf {\bibinfo {volume} {57}},\
  \bibinfo {pages} {4535} (\bibinfo {year} {1998}{\natexlab{a}})},\ \Eprint
  {http://arxiv.org/abs/gr-qc/9701039} {arXiv:gr-qc/9701039 [gr-qc]}
  \BibitemShut {NoStop}%
\bibitem [{\citenamefont {Flanagan}\ and\ \citenamefont
  {Hughes}(1998{\natexlab{b}})}]{Flanagan:1997kp}%
  \BibitemOpen
  \bibfield  {author} {\bibinfo {author} {\bibfnamefont {E.~E.}\ \bibnamefont
  {Flanagan}}\ and\ \bibinfo {author} {\bibfnamefont {S.~A.}\ \bibnamefont
  {Hughes}},\ }\href {\doibase 10.1103/PhysRevD.57.4566} {\bibfield  {journal}
  {\bibinfo  {journal} {Phys. Rev. D}\ }\textbf {\bibinfo {volume} {57}},\
  \bibinfo {pages} {4566} (\bibinfo {year} {1998}{\natexlab{b}})},\ \Eprint
  {http://arxiv.org/abs/gr-qc/9710129} {arXiv:gr-qc/9710129 [gr-qc]}
  \BibitemShut {NoStop}%
\bibitem [{\citenamefont {Droz}\ \emph {et~al.}(1999)\citenamefont {Droz},
  \citenamefont {Knapp}, \citenamefont {Poisson},\ and\ \citenamefont
  {Owen}}]{Droz:1999qx}%
  \BibitemOpen
  \bibfield  {author} {\bibinfo {author} {\bibfnamefont {S.}~\bibnamefont
  {Droz}}, \bibinfo {author} {\bibfnamefont {D.~J.}\ \bibnamefont {Knapp}},
  \bibinfo {author} {\bibfnamefont {E.}~\bibnamefont {Poisson}}, \ and\
  \bibinfo {author} {\bibfnamefont {B.~J.}\ \bibnamefont {Owen}},\ }\href
  {\doibase 10.1103/PhysRevD.59.124016} {\bibfield  {journal} {\bibinfo
  {journal} {Phys. Rev. D}\ }\textbf {\bibinfo {volume} {59}},\ \bibinfo
  {pages} {124016} (\bibinfo {year} {1999})},\ \Eprint
  {http://arxiv.org/abs/gr-qc/9901076} {arXiv:gr-qc/9901076} \BibitemShut
  {NoStop}%
\bibitem [{\citenamefont {Yunes}\ \emph {et~al.}(2009)\citenamefont {Yunes},
  \citenamefont {Arun}, \citenamefont {Berti},\ and\ \citenamefont
  {Will}}]{Yunes:2009yz}%
  \BibitemOpen
  \bibfield  {author} {\bibinfo {author} {\bibfnamefont {N.}~\bibnamefont
  {Yunes}}, \bibinfo {author} {\bibfnamefont {K.~G.}\ \bibnamefont {Arun}},
  \bibinfo {author} {\bibfnamefont {E.}~\bibnamefont {Berti}}, \ and\ \bibinfo
  {author} {\bibfnamefont {C.~M.}\ \bibnamefont {Will}},\ }\href {\doibase
  10.1103/PhysRevD.80.084001} {\bibfield  {journal} {\bibinfo  {journal} {Phys.
  Rev. D}\ }\textbf {\bibinfo {volume} {80}},\ \bibinfo {pages} {084001}
  (\bibinfo {year} {2009})},\ \Eprint {http://arxiv.org/abs/0906.0313}
  {arXiv:0906.0313 [gr-qc]} \BibitemShut {NoStop}%
\bibitem [{\citenamefont {Arun}(2012)}]{Arun:2012hf}%
  \BibitemOpen
  \bibfield  {author} {\bibinfo {author} {\bibfnamefont {K.}~\bibnamefont
  {Arun}},\ }\href {\doibase 10.1088/0264-9381/29/7/075011} {\bibfield
  {journal} {\bibinfo  {journal} {Class. Quantum Grav.}\ }\textbf {\bibinfo
  {volume} {29}},\ \bibinfo {pages} {075011} (\bibinfo {year} {2012})},\
  \Eprint {http://arxiv.org/abs/1202.5911} {arXiv:1202.5911 [gr-qc]}
  \BibitemShut {NoStop}%
\bibitem [{\citenamefont {Chatziioannou}\ \emph {et~al.}(2012)\citenamefont
  {Chatziioannou}, \citenamefont {Yunes},\ and\ \citenamefont
  {Cornish}}]{Chatziioannou:2012rf}%
  \BibitemOpen
  \bibfield  {author} {\bibinfo {author} {\bibfnamefont {K.}~\bibnamefont
  {Chatziioannou}}, \bibinfo {author} {\bibfnamefont {N.}~\bibnamefont
  {Yunes}}, \ and\ \bibinfo {author} {\bibfnamefont {N.}~\bibnamefont
  {Cornish}},\ }\href {\doibase 10.1103/PhysRevD.86.022004} {\bibfield
  {journal} {\bibinfo  {journal} {Phys. Rev. D}\ }\textbf {\bibinfo {volume}
  {86}},\ \bibinfo {pages} {022004} (\bibinfo {year} {2012})},\ \Eprint
  {http://arxiv.org/abs/1204.2585} {arXiv:1204.2585 [gr-qc]} \BibitemShut
  {NoStop}%
\bibitem [{\citenamefont {Berti}\ \emph
  {et~al.}(2005{\natexlab{b}})\citenamefont {Berti}, \citenamefont {Buonanno},\
  and\ \citenamefont {Will}}]{Berti:2004bd}%
  \BibitemOpen
  \bibfield  {author} {\bibinfo {author} {\bibfnamefont {E.}~\bibnamefont
  {Berti}}, \bibinfo {author} {\bibfnamefont {A.}~\bibnamefont {Buonanno}}, \
  and\ \bibinfo {author} {\bibfnamefont {C.~M.}\ \bibnamefont {Will}},\ }\href
  {\doibase 10.1103/PhysRevD.71.084025} {\bibfield  {journal} {\bibinfo
  {journal} {Phys. Rev. D}\ }\textbf {\bibinfo {volume} {71}},\ \bibinfo
  {pages} {084025} (\bibinfo {year} {2005}{\natexlab{b}})},\ \Eprint
  {http://arxiv.org/abs/gr-qc/0411129} {arXiv:gr-qc/0411129} \BibitemShut
  {NoStop}%
\bibitem [{\citenamefont {Yunes}\ and\ \citenamefont
  {Stein}(2011)}]{Yunes:2011we}%
  \BibitemOpen
  \bibfield  {author} {\bibinfo {author} {\bibfnamefont {N.}~\bibnamefont
  {Yunes}}\ and\ \bibinfo {author} {\bibfnamefont {L.~C.}\ \bibnamefont
  {Stein}},\ }\href {\doibase 10.1103/PhysRevD.83.104002} {\bibfield  {journal}
  {\bibinfo  {journal} {Phys. Rev. D}\ }\textbf {\bibinfo {volume} {83}},\
  \bibinfo {pages} {104002} (\bibinfo {year} {2011})},\ \Eprint
  {http://arxiv.org/abs/1101.2921} {arXiv:1101.2921 [gr-qc]} \BibitemShut
  {NoStop}%
\bibitem [{\citenamefont {Alexander}\ and\ \citenamefont
  {Yunes}(2009)}]{Alexander:2009tp}%
  \BibitemOpen
  \bibfield  {author} {\bibinfo {author} {\bibfnamefont {S.}~\bibnamefont
  {Alexander}}\ and\ \bibinfo {author} {\bibfnamefont {N.}~\bibnamefont
  {Yunes}},\ }\href {\doibase 10.1016/j.physrep.2009.07.002} {\bibfield
  {journal} {\bibinfo  {journal} {Phys. Rep.}\ }\textbf {\bibinfo {volume}
  {480}},\ \bibinfo {pages} {1} (\bibinfo {year} {2009})},\ \Eprint
  {http://arxiv.org/abs/0907.2562} {arXiv:0907.2562 [hep-th]} \BibitemShut
  {NoStop}%
\bibitem [{\citenamefont {Yunes}\ and\ \citenamefont
  {Hughes}(2010)}]{Yunes:2010qb}%
  \BibitemOpen
  \bibfield  {author} {\bibinfo {author} {\bibfnamefont {N.}~\bibnamefont
  {Yunes}}\ and\ \bibinfo {author} {\bibfnamefont {S.~A.}\ \bibnamefont
  {Hughes}},\ }\href {\doibase 10.1103/PhysRevD.82.082002} {\bibfield
  {journal} {\bibinfo  {journal} {Phys. Rev. D}\ }\textbf {\bibinfo {volume}
  {82}},\ \bibinfo {pages} {082002} (\bibinfo {year} {2010})},\ \Eprint
  {http://arxiv.org/abs/1007.1995} {arXiv:1007.1995 [gr-qc]} \BibitemShut
  {NoStop}%
\bibitem [{\citenamefont {{Vallisneri}}(2008)}]{2008PhRvD..77d2001V}%
  \BibitemOpen
  \bibfield  {author} {\bibinfo {author} {\bibfnamefont {M.}~\bibnamefont
  {{Vallisneri}}},\ }\href {\doibase 10.1103/PhysRevD.77.042001} {\bibfield
  {journal} {\bibinfo  {journal} {\prd}\ }\textbf {\bibinfo {volume} {77}},\
  \bibinfo {eid} {042001} (\bibinfo {year} {2008})},\ \Eprint
  {http://arxiv.org/abs/arXiv:gr-qc/0703086} {arXiv:gr-qc/0703086} \BibitemShut
  {NoStop}%
\bibitem [{\citenamefont {{Harry}}\ and\ \citenamefont {{LIGO Scientific
  Collaboration}}(2010)}]{2010CQGra..27h4006H}%
  \BibitemOpen
  \bibfield  {author} {\bibinfo {author} {\bibfnamefont {G.~M.}\ \bibnamefont
  {{Harry}}}\ and\ \bibinfo {author} {\bibnamefont {{LIGO Scientific
  Collaboration}}},\ }\href {\doibase 10.1088/0264-9381/27/8/084006} {\bibfield
   {journal} {\bibinfo  {journal} {Class. Quantum Grav.}\ }\textbf {\bibinfo
  {volume} {27}},\ \bibinfo {pages} {084006} (\bibinfo {year}
  {2010})}\BibitemShut {NoStop}%
\bibitem [{\citenamefont {Anderson}\ \emph {et~al.}(2001)\citenamefont
  {Anderson}, \citenamefont {Whelan}, \citenamefont {Brady}, \citenamefont
  {Creighton}, \citenamefont {Chin},\ and\ \citenamefont {Riles}}]{T010110}%
  \BibitemOpen
  \bibfield  {author} {\bibinfo {author} {\bibfnamefont {W.~G.}\ \bibnamefont
  {Anderson}}, \bibinfo {author} {\bibfnamefont {J.~T.}\ \bibnamefont
  {Whelan}}, \bibinfo {author} {\bibfnamefont {P.~R.}\ \bibnamefont {Brady}},
  \bibinfo {author} {\bibfnamefont {J.~D.~E.}\ \bibnamefont {Creighton}},
  \bibinfo {author} {\bibfnamefont {D.}~\bibnamefont {Chin}}, \ and\ \bibinfo
  {author} {\bibfnamefont {K.}~\bibnamefont {Riles}},\ }\href
  {http://dcc.ligo.org} {\emph {\bibinfo {title} {{Beam Pattern Response
  Functions and Times of Arrival for Earthbound Interferometers}}}},\ \bibinfo
  {type} {Tech. Rep.}\ \bibinfo {number} {LIGO-T010110-00-Z}\ (\bibinfo
  {institution} {LIGO},\ \bibinfo {year} {2001})\BibitemShut {NoStop}%
\bibitem [{\citenamefont {{Abadie}}\ \emph {et~al.}(2010)\citenamefont
  {{Abadie}}, \citenamefont {{Abbott}}, \citenamefont {{Abbott}}, \citenamefont
  {{Abernathy}}, \citenamefont {{Accadia}}, \citenamefont {{Acernese}},
  \citenamefont {{Adams}}, \citenamefont {{Adhikari}}, \citenamefont {{Ajith}},
  \citenamefont {{Allen}},\ and\ \citenamefont {et~al.}}]{2010CQGra..27q3001A}%
  \BibitemOpen
  \bibfield  {author} {\bibinfo {author} {\bibfnamefont {J.}~\bibnamefont
  {{Abadie}}}, \bibinfo {author} {\bibfnamefont {B.~P.}\ \bibnamefont
  {{Abbott}}}, \bibinfo {author} {\bibfnamefont {R.}~\bibnamefont {{Abbott}}},
  \bibinfo {author} {\bibfnamefont {M.}~\bibnamefont {{Abernathy}}}, \bibinfo
  {author} {\bibfnamefont {T.}~\bibnamefont {{Accadia}}}, \bibinfo {author}
  {\bibfnamefont {F.}~\bibnamefont {{Acernese}}}, \bibinfo {author}
  {\bibfnamefont {C.}~\bibnamefont {{Adams}}}, \bibinfo {author} {\bibfnamefont
  {R.}~\bibnamefont {{Adhikari}}}, \bibinfo {author} {\bibfnamefont
  {P.}~\bibnamefont {{Ajith}}}, \bibinfo {author} {\bibfnamefont
  {B.}~\bibnamefont {{Allen}}}, \ and\ \bibinfo {author} {\bibnamefont
  {et~al.}},\ }\href {\doibase 10.1088/0264-9381/27/17/173001} {\bibfield
  {journal} {\bibinfo  {journal} {Class. Quantum Grav.}\ }\textbf {\bibinfo
  {volume} {27}},\ \bibinfo {pages} {173001} (\bibinfo {year} {2010})},\
  \Eprint {http://arxiv.org/abs/1003.2480} {arXiv:1003.2480 [astro-ph.HE]}
  \BibitemShut {NoStop}%
\end{thebibliography}%
\end{document}